\pdfoutput=1
\documentclass[12pt,preprint]{aastex}

\usepackage{epstopdf}
\usepackage{array}

\usepackage{longtable}  \usepackage{lscape} 
\usepackage{gensymb}

\shorttitle{XRB Catalog}
\shortauthors{P. Jenke et al.}

\begin{document}


\title{The Fermi-GBM 3-year X-ray Burst Catalog}


\author{P. A. Jenke\altaffilmark{1}, M. Linares\altaffilmark{2,3,4.9}, V. Connaughton\altaffilmark{5}, E. Beklen\altaffilmark{6,10}, A. Camero-Arranz\altaffilmark{7,5}, M. H. Finger\altaffilmark{5}, C. A. Wilson-Hodge\altaffilmark{8}}

\altaffiltext{1}{CSPAR, SPA University of Alabama in Huntsville, Huntsville, AL 35805, USA}
\altaffiltext{2}{Instituto de Astrof{\'i}sica de Canarias, c/ V{\'i}a L{\'a}ctea s/n, E-38205 La Laguna, Tenerife, Spain}
\altaffiltext{3}{Universidad de La Laguna, Dept. Astrof{\'i}sica, E-38206 La Laguna, Tenerife, Spain}
\altaffiltext{4}{Institutt for fysikk, NTNU, Trondheim, Norway}
\altaffiltext{5}{Universities Space Research Association, Huntsville, AL 35805, USA.}
\altaffiltext{6}{Department of Physics, Suleyman Demirel University, 32260, Isparta, Turkey}
\altaffiltext{7}{Institut de Ci\`{e}ncies de l'Espai, (IEEC-CSIC), Campus UAB, Fac. de Ci\`{e}ncies, Torre C5 pa., 08193, Barcelona, Spain}
\altaffiltext{8}{Marshall Space Flight Center, Huntsville, AL 35812, USA}
\altaffiltext{9}{MIT Kavli Institute for Astrophysics and Space Research, Massachusetts Institute of Technology, 70 Vassar Street, Cambridge, MA 02139, USA}
\altaffiltext{10}{Department of Physics and Astronomy, West Virginia University, P.O. Box 6315, Morgantown, WV 26506, USA}


\begin{abstract}
The Fermi Gamma Ray Burst Monitor (GBM) is an all sky gamma-ray monitor well known in the gamma-ray burst community.  Although GBM excels in detecting the hard, bright extragalactic GRBs, its sensitivity above 8 keV and all-sky view make it an excellent instrument for the detection of rare, short-lived Galactic transients.  In March 2010, we initiated a systematic search for transients using GBM data.  We conclude this phase of the search by presenting a 3 year catalog of 1084 X-ray bursts.  Using spectral analysis, location and spatial distributions we classified the 1084 events into 752 thermonuclear X-ray bursts, 267 transient events from accretion flares and X-ray pulses, and 65 untriggered gamma-ray bursts.  All thermonuclear bursts have peak blackbody temperatures broadly consistent with photospheric radius expanison (PRE) bursts. We find an average rate of 1.4 PRE bursts per day, integrated over all Galactic bursters within about 10 kpc. These include 33 and 10 bursts from the ultra-compact X-ray binaries 4U~0614+09 and 2S~0918-549, respectively. We discuss these recurrence times and estimate the total mass ejected by PRE bursts in our Galaxy.\end{abstract}



\section{Introduction}
\label{sec:intro}

Rare, unpredictable and transient astronomical phenomena are 
difficult to observe due to their very own nature, yet they often 
lead to exciting astrophysical discoveries.
At any wavelength, the most efficient way of detecting rare transients 
is to maximize the observed field of view (FoV).
The high-energy (X-ray and Gamma-ray) sky can vary rapidly, on 
timescales much shorter than hours.
If we are interested in {\it short-lived} rare transient phenomena (seconds 
to minutes long), the most relevant observational capability is the 
{\it instantaneous} FoV.
Even though serendipitous detections occur, neither pointed narrow FoV 
instruments nor all-sky monitors based on scanning techniques are well 
suited to catch such short and rare events.

The Gamma-ray burst monitor (GBM) onboard the {\it Fermi} observatory
has an instantaneous FoV of about 75\% of the sky \citep{Meegan2009} and is sensitive to photon energies down to 8~keV.
%
%
Even though it was designed to detect and characterize Gamma-ray bursts 
(GRBs), these characteristics make GBM a unique instrument to detect
rare, short and bright X-ray bursts (XRBs).
In March 2010, we initiated a systematic search for XRBs using
{\it Fermi}-GBM data (Section~\ref{sec:GBM-XBM}). 
In the first three years, this search has yielded 752 thermonuclear
X-ray bursts (tXRBs; Secs.~\ref{sec:intro-tXRBs} \& \ref{sec:tXRB}), 267
transient events from accretion flares and X-ray pulses (aFXPs), as
well as 65 untriggered long gamma ray bursts (uGRBs).
We present here the {\it Fermi}-GBM 3-year X-ray Burst Catalog and
summarize its main results with an emphasis on tXRBs.

\subsection{The rare and most energetic thermonuclear bursts}
\label{sec:intro-tXRBs}

The accreted matter in neutron star low-mass X-ray binaries (NS-LMXBs)
piles up on the neutron star surface, reaching regions of increased
density and becoming fuel for thermonuclear reactions.
When ignition conditions are met at the bottom of the accreted shell,
unstable reactions trigger a thermonuclear runaway that quickly burns
the pile of fuel, generally a mix of hydrogen (H), helium (He) and
heavier elements (``metals'').
This cyclic phenomenon has been observed for four decades in what is
know as thermonuclear (type I X-ray) bursts \citep{Grindlay76,Belian76}.

The main parameter which sets the frequency or {\it recurrence time}
of thermonuclear X-ray Bursts (tXRBs) is the {\it mass accretion rate}
per unit area, $\dot{m}$ \citep{Fujimoto81,Bildsten98}.
The main reason is simple: $\dot{m}$ sets the rate at which fuel is
replenished between tXRBs.
However, other factors play an important role in the tXRB recurrence
time, including composition and the thermal state of the NS envelope.
In particular, at the lowest $\dot{m}$ (near or below 1\% of the
Eddington limit) the heat flux from the NS crust can critically
influence the ignition conditions for tXRBs.
Thus we can potentially use low-$\dot{m}$ tXRBs to constrain the
internal properties of NSs \citep{Cumming06}.
However, because recurrence times of low-$\dot{m}$ tXRBS are of the
order of weeks to months, they are extremely difficult to measure with
pointed or scanning X-ray detectors.
GBM has opened a new window to these
events, and it is yielding the first accurate measurements of their
recurrence times \citep{Linares12b}.

\section{The Fermi-GBM X-ray Burst Monitor}
\label{sec:GBM-XBM}

	 GBM is an all sky monitor whose primary objective is to extend the energy range over which gamma-ray bursts are observed in the Large Area Telescope (LAT) on \emph{Fermi} \citep{Meegan2009}.  GBM consists of 12 NaI detectors with a diameter of 12.7 cm and a thickness of 1.27 cm and two BGO detectors with a diameter and thickness of 12.7 cm.  The NaI detectors have an energy range from 8 keV to 1 MeV while the BGOs extend the energy range to 40 MeV.  The GBM flight software was designed so that GBM can trigger on-board in
response to impulsive events, when the count rates recorded in two or more NaI detectors significantly exceed the background count rate on at least one time-scale from 16 ms to
4.096 s in at least one of four energy ranges above 25 keV. The lower energy and longer
time-scales are inaccessible to the on-board triggering algorithms owing to strong variations
in background rates that are incompatible with a simple background modeling needed for
automated operation on a spacecraft.  Between 25 and 50 keV, only the shortest time-scales
are probed on-board (under 128 ms). We report here on our search of GBM continuous data for impulsive events that are too long and too spectrally soft to trigger on-board.    

GBM has three continuous data types: CTIME data with nominal 0.256-second time resolution and 8-channel spectral resolution used for event detection and localization, CSPEC data with nominal 4.096-second time resolution  and 128-channel spectral resolution which is used for spectral modeling, and Continuous Time Tagged Event (CTTE) data with  time stamps ($2 \mu$s precision) on individual events at full 128-channel spectral
resolution that was made available November 2012.  The NaI CTIME and CSPEC data from 8-50 keV are used in the following analysis.
	 
%

 \subsection{Data Selection}
The Fermi-GBM X-ray Burst Monitor relies on daily inspection of CTIME channel 1 (12-25 keV) data and began operations in 2010 March 12.  The CTIME data are rebinned to a minimum of 0.25 second time bins to adjust intervals of high resolution data initiated by instrument triggers.  NaI detector rates, from all 12 detectors and channels 0-2 (8-50 keV), are automatically filtered  removing phosphorescence events, times of high total rates, 
times near the SAA and intervals of rapid spacecraft slews. An empirical background model is
fit to the detector rates in each channel (0-2) and each detector. The background model has terms to account for
bright sources and their Earth occultations plus a quadratic spline model to account for the low frequency trends of the remaining background (below $\sim$1 mHz).  The background model is visually compared to the rates in the energy band between 12 and 25 keV with a time resolution of $\sim8.2$ seconds.   Transient events that rise above the background model are saved by manually selecting the corresponding time intervals.  Source rates and background rates for the first three energy bands (8 - 50 keV) along with mid-times of these 
manually-selected time intervals are recorded.  Between March 2010 and March 2013, the search resulted in 5093 selected events.

	type I X-ray bursts, the softest population of events likely to be detected, are expected to have a blackbody spectrum with a temperature between about 0.5 and 3 keV.  Due to the gradual rollover in the expected photon spectrum between 12 and 25 keV and the steep drop in effective area in CTIME channel 0 ($\sim8$-12 keV) data \citep{Meegan2009}, channel 1 (12-25 keV) is the most sensitive channel to these XRBs.  The choice of 8.2 second timing resolution for channel 1 data 
is a compromise between the desire to maximize our sensitivity to these events and the time demands of this labor-intensive process,
and limits the minimal detectable burst duration to around 10 seconds.  Background count rate variations 
over the $\it Fermi$ orbit, caused both by changes in geomagnetic latitude and varying spacecraft attitude,  prevent visual identification of very long bursts.   Our search is thus sensitive to bursts and flares with durations in the 10 -- 1000 s range.

\subsection{Localization}
\label{sec:localization}
Localization of our events of interest utilizes the angular response of the NaI detectors
to reconstruct the most likely
arrival direction based on the differences in background-subtracted count rates recorded in 12 NaI detectors that have different sky orientations.
The method is adapted from the method used for GBM GRB localization \citep{Connaughton_2015}, with a
cruder background fitting method. We use data between 12 -- 50 keV
and the model rates more suitable for sources with softer energy spectra: 
galactic transients (power-law with index = -2),  solar flares (power-law with index = -3), and type I XRBs (blackbody with temperature = 4 keV).
This process yields a localization and a 68\% statistical uncertainty radius (assuming a circular uncertainty region), $\sigma$.
We also determine a goodness-of-fit parameter, $\chi^{2}$, of the localization. Other parameters of interest include a rough event duration,  
a list of detectors with an angle between source and detector normal less than $60^\circ$, the net count rates in these detectors, 
and hardness ratios derived from count rates in different energy channels. 
If the event localizes to within $10^\circ$ of the centroid of the solar disk or is less than 3$\sigma$ from the Sun position then the event is rejected, as are
events with localizations clearly (beyond the statistical uncertainty) beneath the Earth's horizon.  

If the net count rates of the two brightest detectors are inconsistent with a single source direction then the event is rejected.  Such events may occur during a particle shower within or near the spacecraft and are not associated with an astronomical source.  An additional check is performed to eliminate particle events which originate in the spacecraft.  These events appear to have a hard spectra and thus might be initially classified as uGRBs but unlike GRBs their light curves in the 50-300 keV range are very similar for all 12 NaI detectors.  This produces a poor $\chi^{2}$ in the localization fit and we use a cut-off in $\chi^{2}$ of 1000 to reject these particle events, more tolerant than reported
in \citep{Connaughton_2015} because the quality of the background fits over the low-energy channel data analyzed here is more variable, and even real astrophysical
events may produce localizations with large $\chi^2$ values.  All other events are considered XRB candidates.  Once the events are localized, they are searched for temporal and spatial coincidence with GBM and \emph{Swift} triggered GRBs.  If the XRB candidate event locates within 3$\sigma$ of a triggered GRB and the XRB candidate event mid-time occurs within 150\% of the T-90 duration of the the GRB trigger time then the XRB event is considered a triggered GRB and rejected.  After these filtering steps there are 2253 events remaining of the original manually selected sample of 5093.  The vast majority of rejected events were identified as solar flares. 

\subsection{Spectral Analysis}  
\label{sec:spetral}
Response matrices for each XRB candidate event are created from a response model constructed from simulations incorporating the Fermi spacecraft mass model into GEANT4 (\cite{GEANT4}).  CSPEC data are used for spectral analysis in RMFIT, a forward-folding spectral analysis software often used in GBM gamma ray burst studies.\footnote{https://gamma-wiki.mpe.mpg.de/GBM/RMFITPublicReleasePage}   Through localization and visual inspection (see Section \ref{sec:aFXP}) many of the events were identified with Sco X-1 and Vela X-1 (aFXPs) and spectral analysis was not necessary for identification.  We did; however, performed spectral analysis on a few of these in order to aid in the association of those events in which identification was not apparent.  Blackbody and power-law models are fit to all of the remaining data, the former because it is physically motivated for tXRBs and the latter because it is a simple model that can be used to fit a variety of events, and may be useful to classify their spectral hardness even if the model does not fully describe the data.  

In the course of our spectral analysis, we identified fits for which
the residuals of the unfolded spectrum for different detectors were inconsistent with each other.
This is evidence for a bad localization which we attributed to poor background fits in one or more detectors.  We selected background time intervals before and after the source time interval, as is done for GRB localization by the GBM team.   We fit the selections with a polynomial (usually a quadratic but occasionally a higher order polynomial is necessary to fit the data) for each detector and for each channel between 0 and 2 (8-50 keV).  The event is selected and the fitted backgrounds subtracted.  A new localization is performed and the event is labeled as before.  Subsequent localizations almost invariably provided an improved localization $\chi^{2}$ and smaller error.  This was most often due to the previous background fit including the source and reducing the residual rates in each detector in a non-uniform manner thus producing an erroneous localization and resulting in poor detector responses.  

\begin{figure}[h!]
\centering
  \begin{tabular}{c}
   \includegraphics[width=4in]{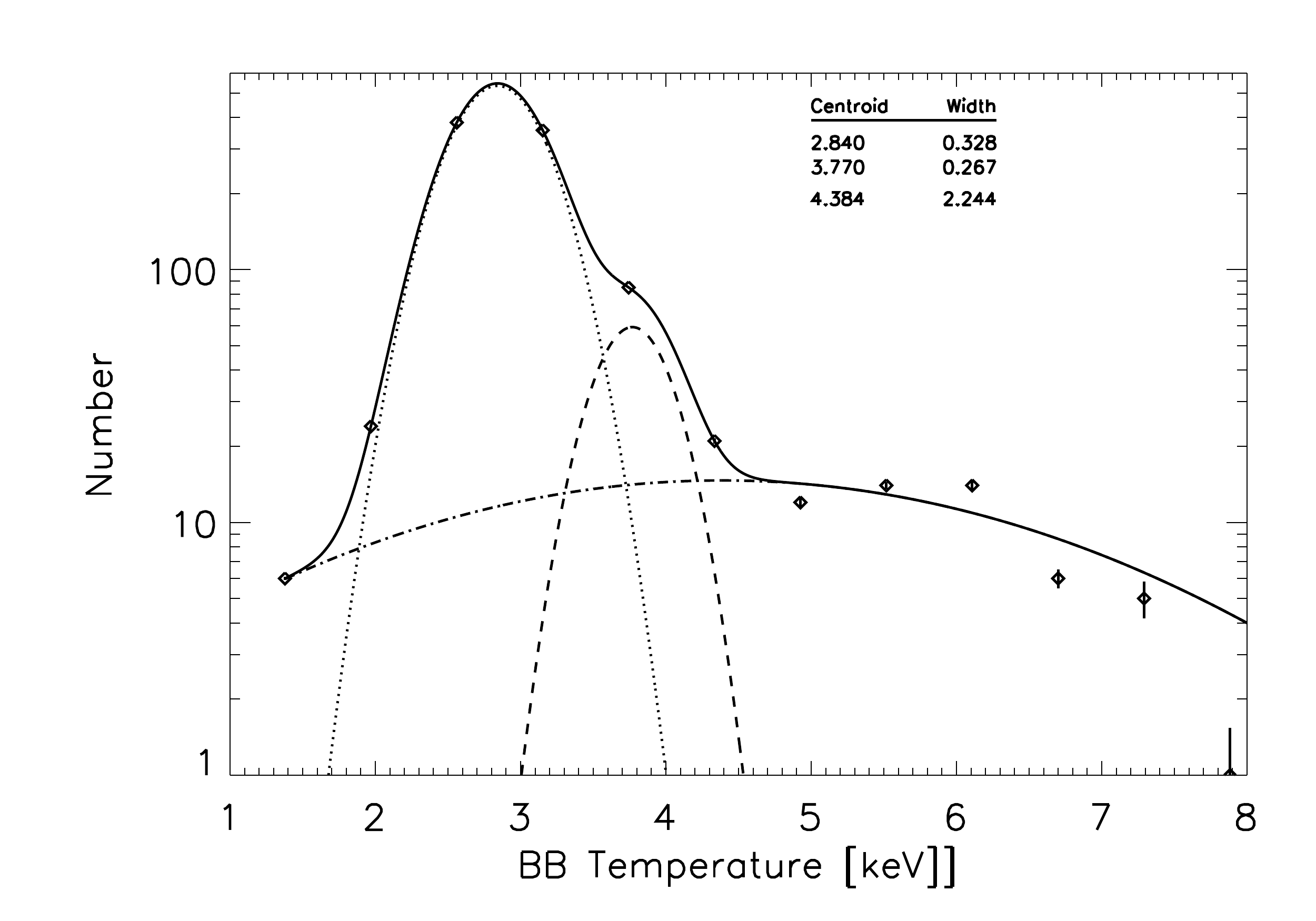} \\
   \includegraphics[width=4in]{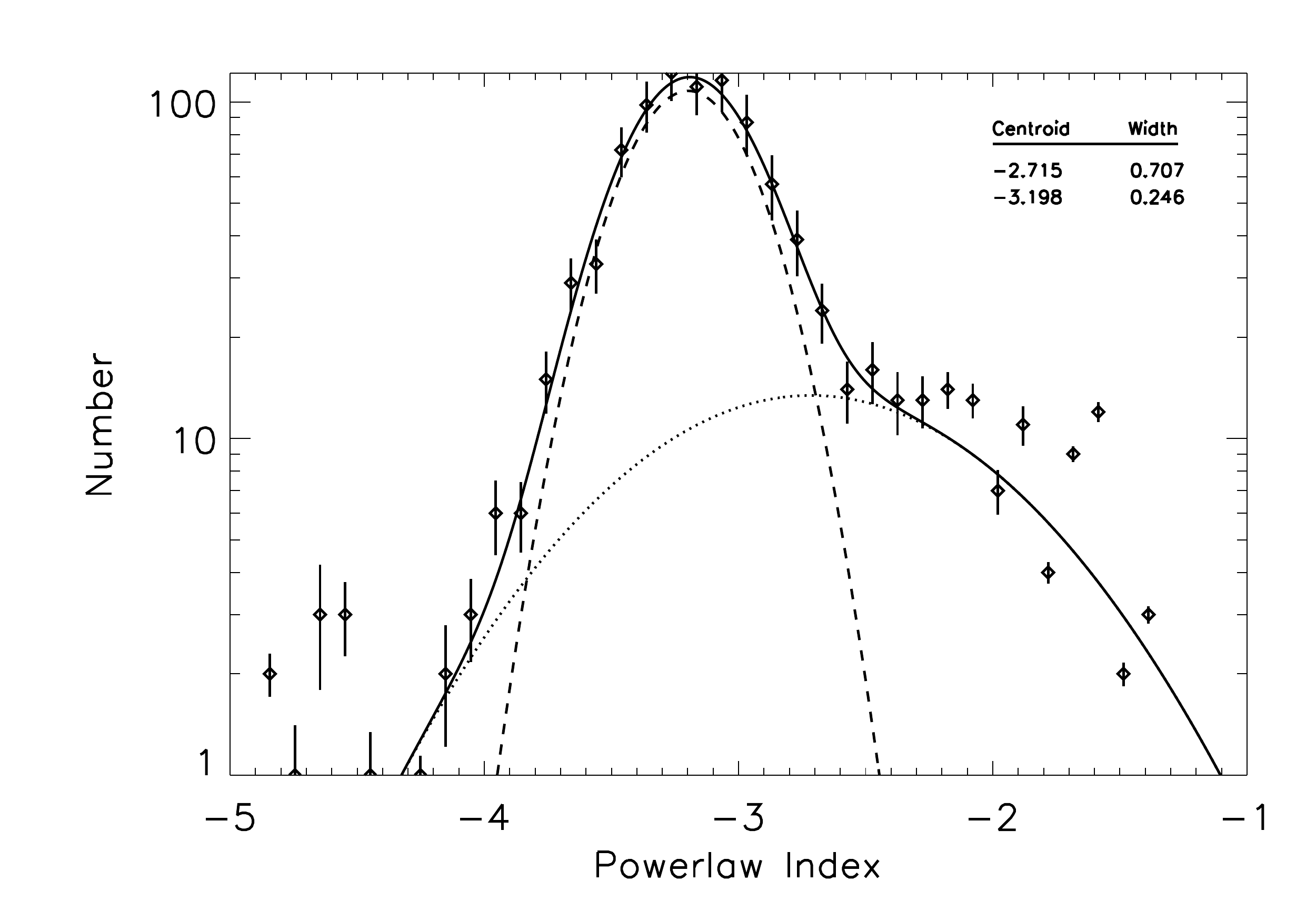} 
   \end{tabular}
   \caption{Top: Distribution of the temperatures in keV from the blackbody fits to the XRB event spectra.  Diamonds are the data points.  The solid line is the total model and the dashed and dotted lines are the gaussian components of the total model.  Bottom: Distribution of the index from the power-law fits to the XRB event spectra.  Diamonds are the data points.  The solid line is the total model and the dashed and dotted lines are the gaussian components of the total model.}
  \label{fig:Spectrum}
 \end{figure}

Weak events (85) in which spectral analysis was not possible were rejected.  With these rejected events and events that were reclassified as solar due to the new localization, there remained 1084 XRB candidates.  Figure \ref{fig:Spectrum}  shows the results of the spectral modeling of these XRB candidates.  The top panel is a histogram of the resulting temperatures from blackbody fits while the lower panel is a histogram of the indices from power-law fits.  The histograms are fit to a model that consists of multiple gaussians.  The power-law distribution is well fit with two gaussians with $\chi^{2} = 85$ with 101 DOF.  The blackbody distribution required three gaussians for a fit with $\chi^{2} = 87$ with 56 DOF.

\begin{figure}[h!]
\centering
  \begin{tabular}{c}
   \includegraphics[width=4in]{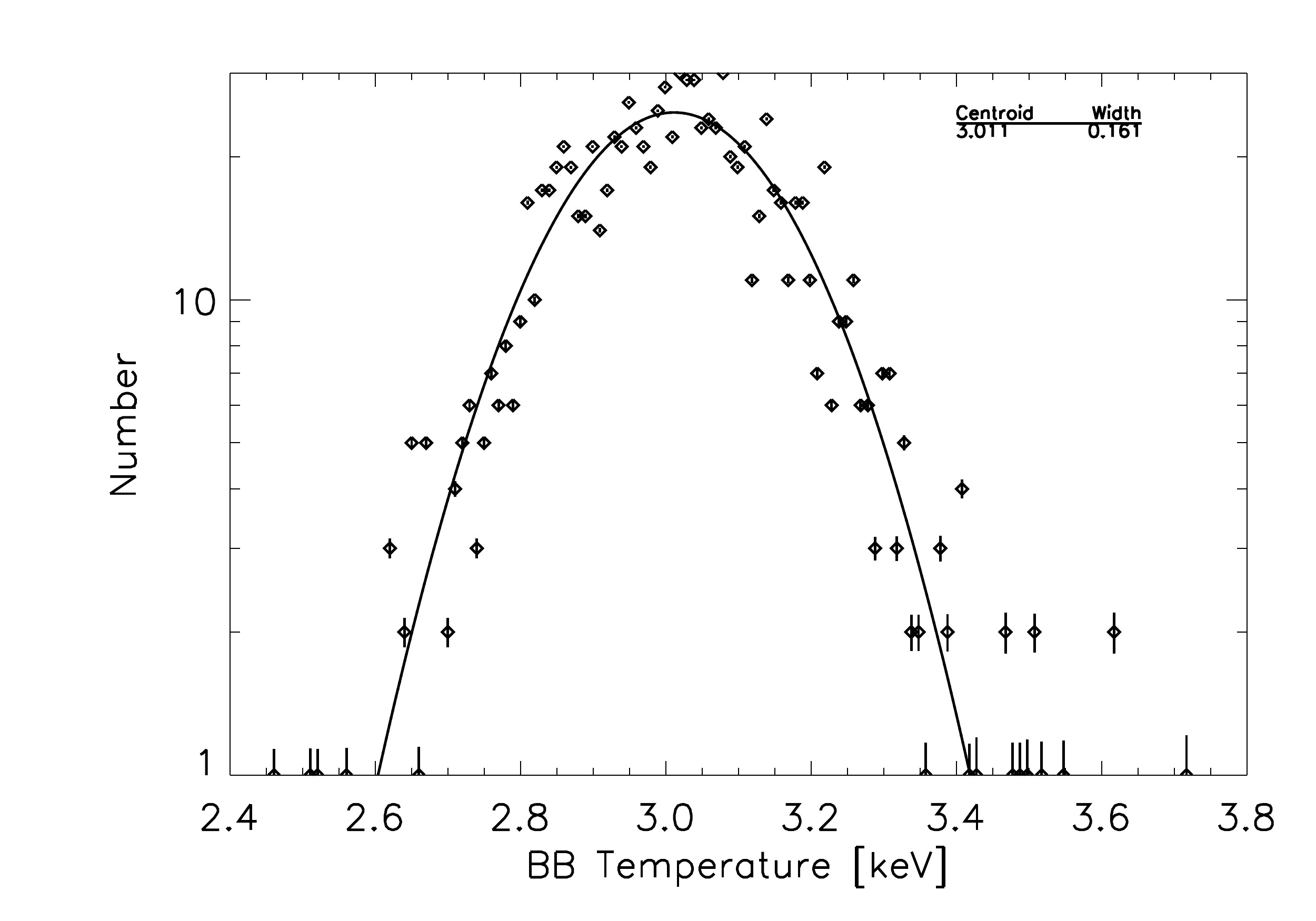} \\
   \includegraphics[width=4in]{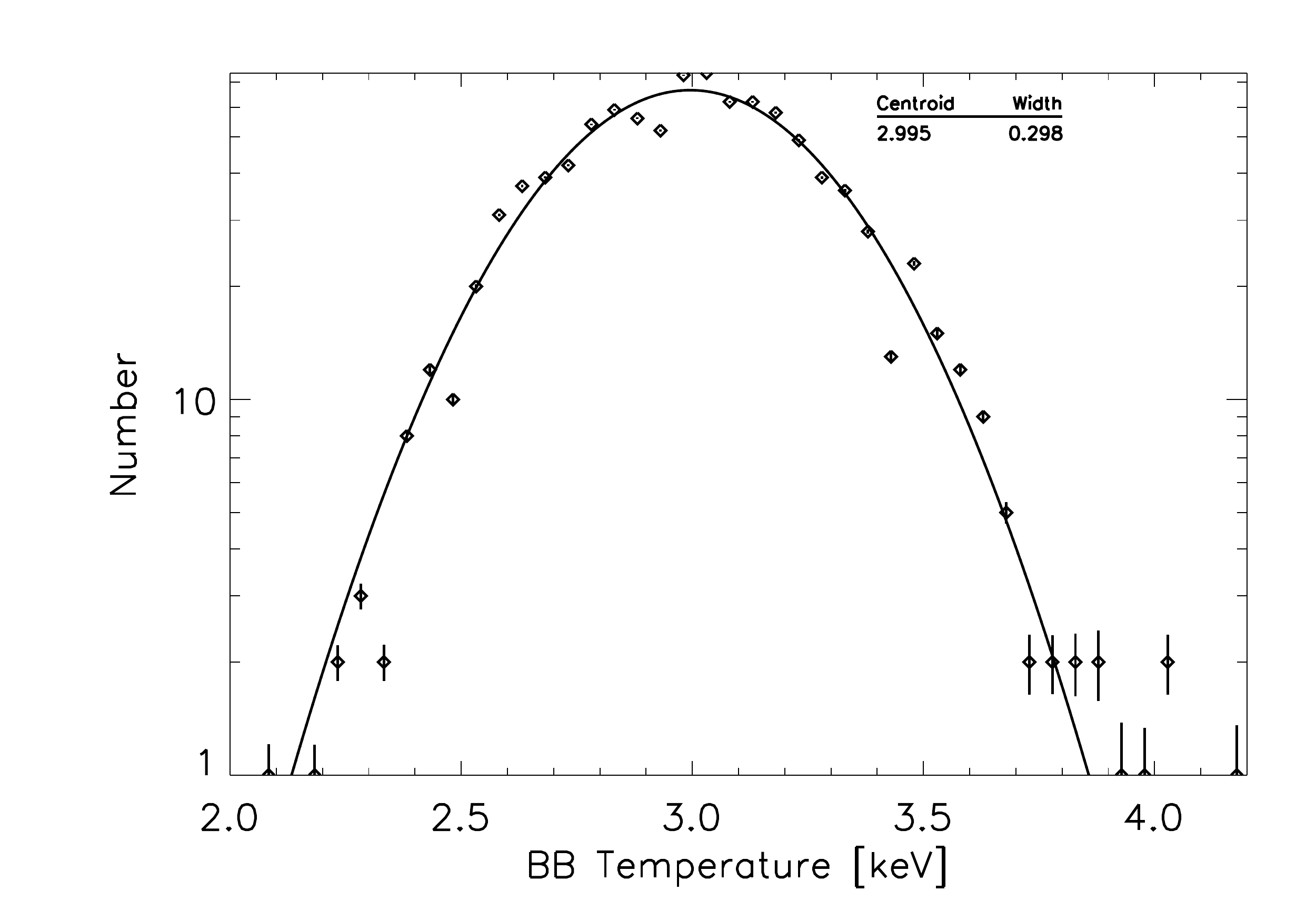} 
   \end{tabular}
   \caption{Top: Distribution of the temperatures in keV from the blackbody fits to the simulated data from the peak of event 10032800979.  Diamonds are the data points.  The solid line is the best fit gaussian.  Bottom: Distribution of the temperatures in keV from the blackbody fits to the simulated data from the tail of event 10032800979.  Diamonds are the data points.  The solid line is the best fit gaussian.}
  \label{fig:sim}
 \end{figure}

 The energy spectra of tXRBs is expected to be a blackbody with temperature between 0.5 and 3.0 keV.  Our results for 4U 0614+09 \citep{Linares12b} and Figure \ref{fig:Spectrum} suggest that GBM is sensitive to bursts with
temperatures at the high end of this range.
The characteristics of event 10032800979 (Table~\ref{tbl:type1_table}), which has been identified as a tXRB from 4U 0614+09 \citep{Linares12b}, was used to demonstrate our ability to  recover 
the temperature using the spectral analysis approach described above.   
1000 simulated data sets were created for the NaI detectors N0, N1, N9 and NA for the brightest 4.1 second bin which has an energy flux (10-100 keV) of ($7.57 \pm 0.33$)E$^{-8}$ erg s$^{-1}$ cm$^{-2}$ and another 1000 data sets for a 4.1 second bin in the tail of the burst which has an 10 -100 keV energy flux of ($3.12 \pm 0.29$)E$^{-8}$ erg s$^{-1}$ cm$^{-2}$.  
The best fit temperatures for this burst in these time bins are $3.28 \pm 0.12$ keV and $3.00 \pm 0.24$ keV respectively.   A blackbody spectrum with a temperature of 3.0 keV is used to simulate the data.  The simulated data are fit to a blackbody spectrum resulting in the best fit spectral temperature centered on $3.0 \pm 0.16$ keV for the brightest interval and $3.0 \pm 0.3$ keV for the weak interval assuming the temperatures are normally distributed.   The resulting temperature distributions and fits to a gaussian function are shown in Figure \ref{fig:sim}.
These results indicate that any systematic error in the spectral analysis is not dominated by the lack of spectral sensitivity in GBM at these energies and fluxes.  

We checked if scattering off the Earth's limb was a possible contributor to the systematic error in the spectral analysis by checking the proximity of the events to the Earth's limb.  There were only 23 events that were within 100 seconds of the Earth's limb and only 2 less than 20 seconds.  These last two events had a blackbody temperature of $3.3 \pm 0.2$ keV and  $2.9 \pm 0.2$ keV and were within 10 degrees of the Galactic center.  We do not expect such limb events to be a source of systematic error in our catalog.

\subsection{Temporal Analysis}
\label{sec:duration}
Temporal analysis of XRB events include the calculation of event duration, rise times, and decay times and was performed after classification (see Section \ref{sec:catalog}) was finished. Due to the nature of the aFXPs (see Section \ref{sec:aFXP}), these events were excluded from the temporal analysis.  Durations for these events are taken from the time interval of the original event selection.  
Since this analysis requires detailed visual inspection of the light curve, these events underwent additional scrutiny to ensure that aFXPs did not contaminate the remains categories.   

For each event where durations are calculated, light curves for all detectors are visually inspected in the 12-25 keV energy band and background regions are selected.  The background is fit to a polynomial (usually a quadratic but occasionally a higher order polynomial is necessary to fit the data).  The background fit is then subtracted from the light curve.  The detectors, in which signal is evident, are selected and the first three energy channels (8-50 keV) are added together and displayed as a single light curve.  The peak intensity of the light curve (t$_{peak}$) is selected.   The times at 25\% of peak during the rise (t$_{25}$), 90\% of peak during the rise (t$_{90}$) and 10\% of the peak along the decay (t$_{10}$) are calculated.   As in \cite{Galloway_2008}, the rise time $t_{rise}$ is the time for the intensity to rise from 25\% to 90\% of its peak value, the duration of the event is defined as  t$_{10}$ - t$_{25}$ and the decay time (t$_{decay}$) is defined as (t$_{10}$ - t$_{peak}$).  

\begin{figure}[h!]
\centering
   \includegraphics[width=4in]{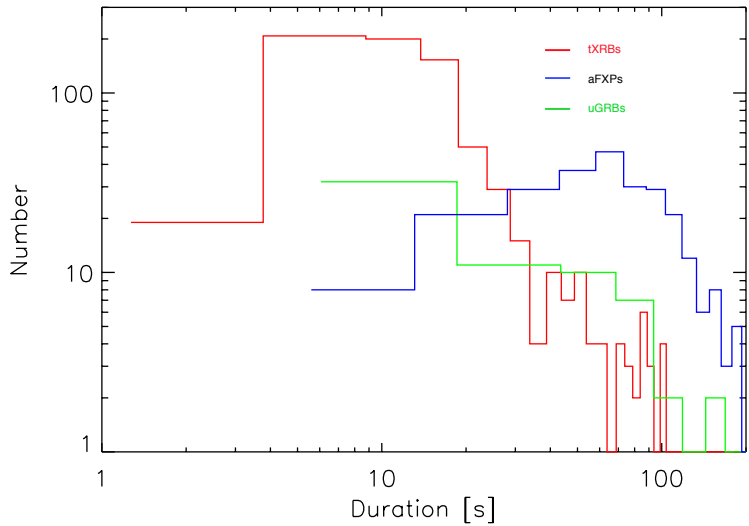}
  
    \caption{Duration distribution for the three categories of events in the catalog.  The red curve is the duration distribution for the tXRBs, the green curve is the duration distribution for the uGRBs, and the blue curve is the duration distribution for the aFXPs.  }
   
  \label{all_duration}
 \end{figure}

Figure \ref{all_duration}  shows the duration distribution for the three categories separated by color.  The durations for each category show considerable overlap and are not used to distinguish between categories.

\section{X-ray Burst Catalog}
\label{sec:catalog}
 The three year XRB catalog contains 1084 events occurring between MJD 55267 and 56347 (2010 March 12 - 2013 February 24) which are classified into three categories: the tXRBs, the aFXPs,  and the uGRBs.   Clear distinctions between the three categories is not possible; therefore, we make the following quantitative effort.  First, the aFXPs are categorized based on location, visual inspection of the light curve, and spectral analysis (see Section \ref{sec:aFXP}).  Second, the tXRBs are categorized using spectral analysis alone and then the uGRB events are categorized based on spectral analysis and location.
 
\begin{figure}[h!]
 \centering
   \includegraphics[width=4in]{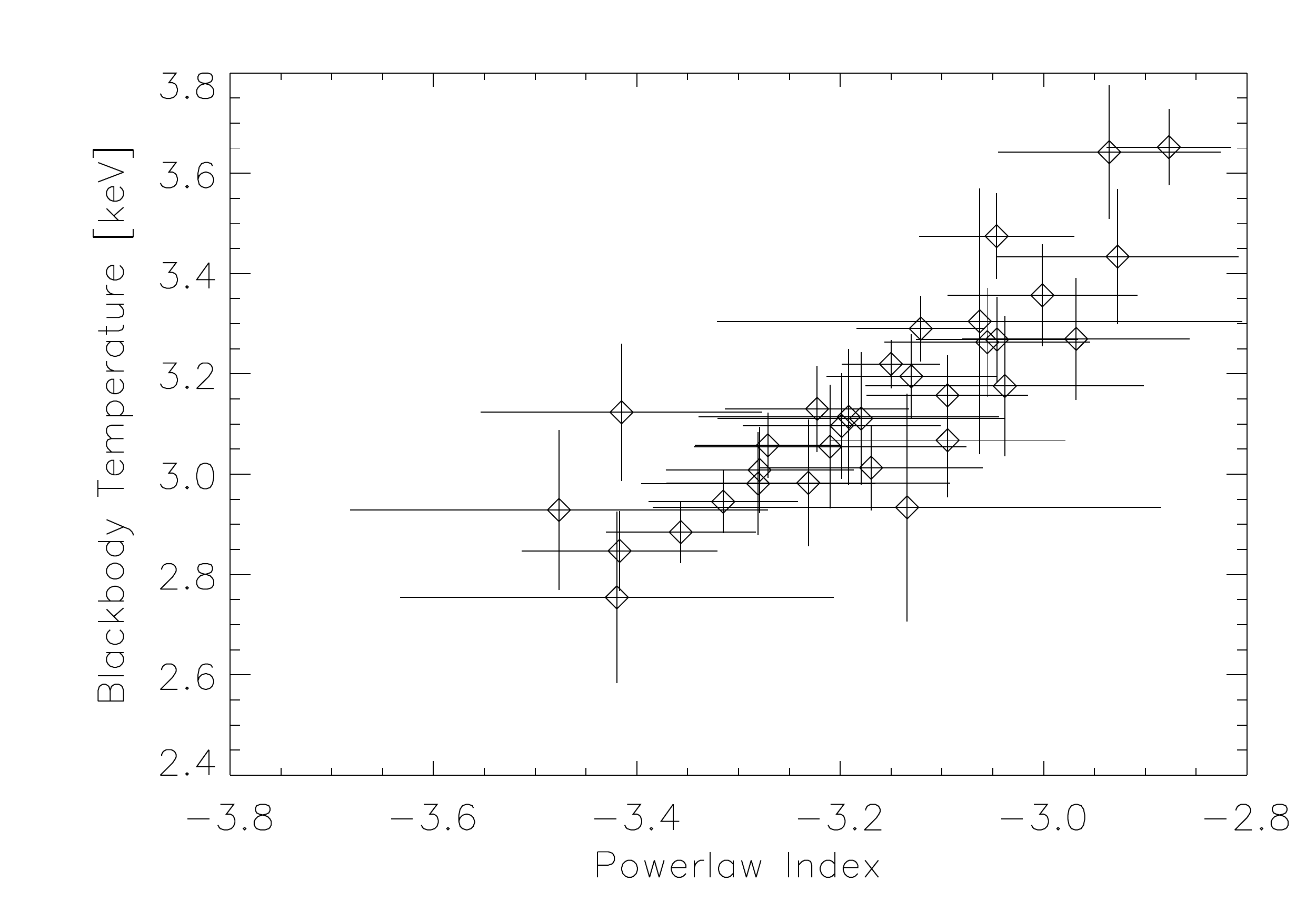}
\caption{Blackbody temperature distribution for the XRB candidates associated by location and spectral shape with 4U 0614+09 show
a close correlation with the index of the power-law fits to the same events.}
   \label{fig:bb_pl}
\end{figure}

 The XRB events are from a wide variety of sources and their spectra is expected to be just as varied.  The tXRBs are expected to have a blackbody spectrum (0.5 - 3 keV) while many of the aFXPs and uGRBs are expected to have non-thermal specta which may be modeled, in part, by a power-law.  Although the power-law spectral model is generally not a good choice for all three categories of events, it serves well as an indicator of spectral properties for which all categories may be compared.  We used spectral results from 32 events that we confidently associate with 4U 0614+09 from this work and \cite{Linares12b} to compare the spectral fit results from a blackbody and power-law model (see Figure \ref{fig:bb_pl}).  There is a tight correlation between the blackbody temperature and the index from a power-law fit justifying our sole use of the power-law in spectral comparisons.  This correlation, when considering all events, is tight up to 4 keV (index = -2.5) after which there is considerable scatter in the blackbody temperature.   

\begin{figure}[h!]
\centering
    \includegraphics[width=3in]{pl_dist.eps} 
    \includegraphics[width=2.5in]{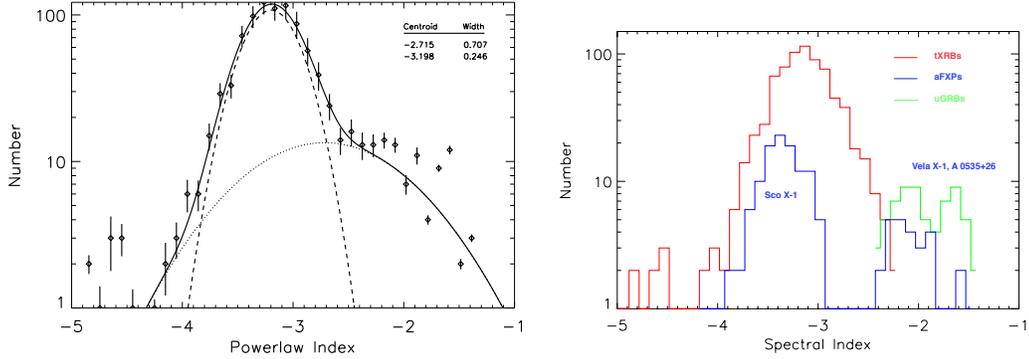} 
   \caption{Distribution of the spectral index from power-law fits to the XRB candidate spectra.  
Left: Diamonds are the data points for all the XRB events.  
The solid line is a model fit to the data and the dashed and dotted lines are the two gaussian components of the total model.
Right: Separation of indices by class of event. 
The red curve is the index distribution for the tXRBs.  The blue curve is the index distribution for the aFXPs while the green curve is the index distribution for the uGRBs.  Contributions from Sco X-1 and Vela X-1 (both aFXPs) are marked.}
  \label{all_spectra}
 \end{figure}

We choose to be inclusive with our category of tXRBs and use a cut-off in spectral index of -2.5 (4 keV).  Any event which is not an aFXP and has a spectral index that is consistent (1$\sigma$) with being softer than -2.5 ($<4$ keV) is categorized as a tXRB (See the red distribution in the right panel of Figure \ref{all_spectra}).  This exceeds, by a good margin, the theoretical maximum temperature for type I bursts (\cite{Boutloukos_2010}).
The left panel of Figure \ref{all_spectra} shows the distribution of power-law indices for all XRB events.  The softer distribution has a centroid of $-3.2\pm0.25$ and, being the softer distribution, is expected to contain the tXRBs.  The spectral index cut-off of -2.5 represents a $3\sigma$ departure from the centroid thus validating our choice. 

The uGRBs are expected to be isotropically distributed across the sky while the tXRBs are mostly at the Galactic center.   If we assume GBM uniform exposure, the power-law index cutoff that maximizes the source distribution isotropy can also be used to distinguish these two categories.  Using the Rayleigh Test, the maximum isotropy ($\chi^2 = 1.8/3$ dof) occurs for those events whose power-law index is consistent ($1 \sigma$) with being greater than -2.43 thus again validating our choice of -2.5 as a spectral index discriminator between the uGRBs and the tXRBs.   Three uGRB events had a spectral index between -2.43 and -2.5 and could arguably be placed in the tXRB classification and they were 10101041428, 11100350666, and 12062078172.

  \begin{figure}[h!]
\centering
    \includegraphics[width=4in]{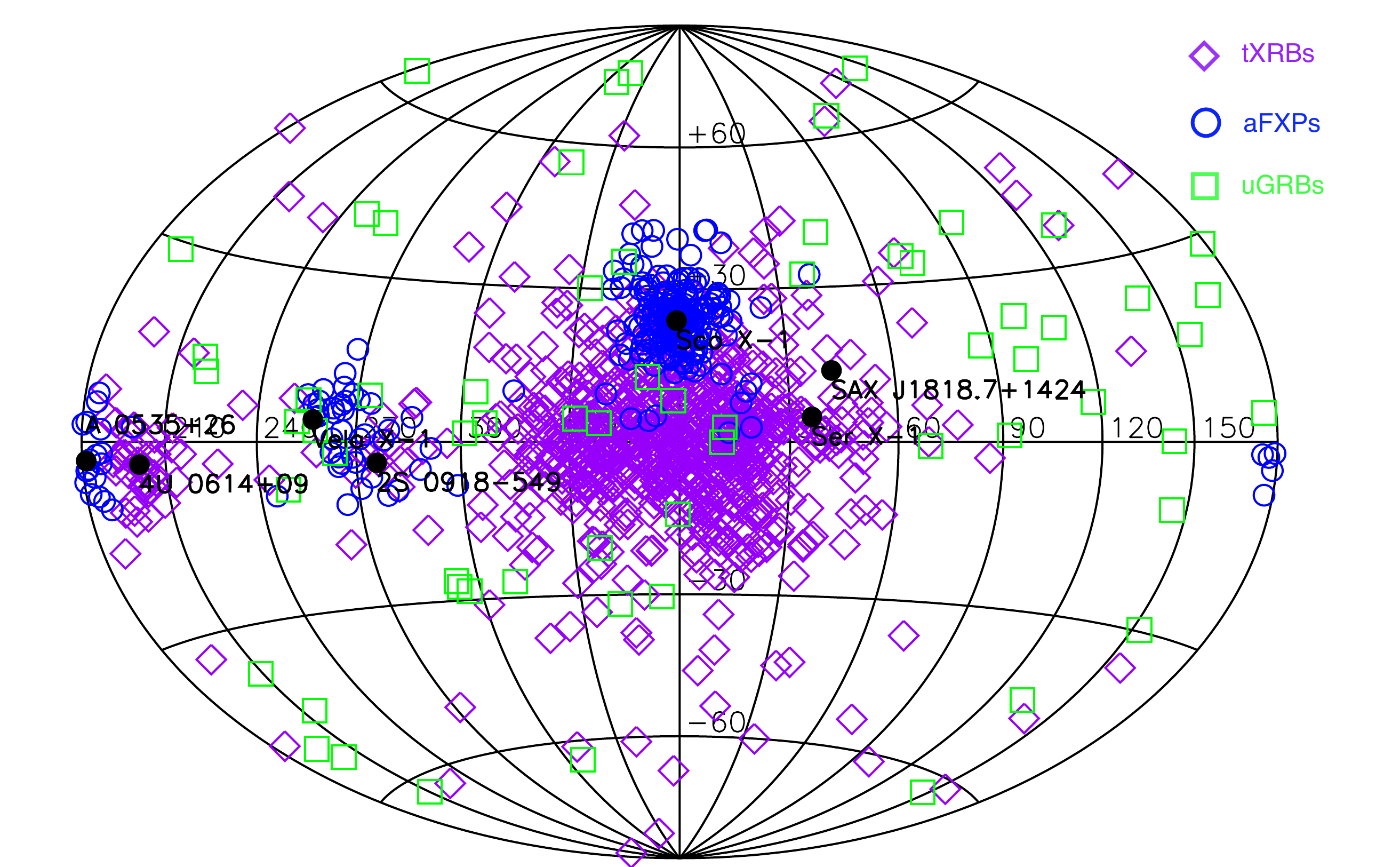} 
   \caption{Centroids of the localization of all events in Galactic coordinates.  The purple diamonds are the locations of the tXRBs.  The blue circles are the location of the aFXPs, the green squares are the uGRBs.  The error circles for the localization are generally larger than their symbols.}
  \label{galactic_dist}
 \end{figure}

  Figure \ref{galactic_dist} shows the location of all the events in Galactic coordinates with the categories distinguished by color and symbol.  The purple diamonds are the tXRBs and there is a large number distributed around the Galactic center which is consistent with the distribution of the known type I XRBs.  There is a smaller cluster of events consistent with the location of 4U 0614+09.  The aFXPs are shown as blue circles and are largely in three clusters centered around A 0535+26, Vela X-1 and Sco X-1.  The green squares are the uGRBs which are distinguished by their isotropic distribution.  The classification scheme is summarized in table \ref{tbl:totals}.
 \begin{table*}[h!]
 \caption{Source Classification Summary} 
\begin{tabular}{lccc}
\tableline\tableline
Category & Number Events & Selection & Properties  \\
                   &      Events          & Process    & d\\ 
\tableline
aFXPs  & 267 & Location, Visual Inspection & Periodic, Continuous flares \\
             &        &  Spectral: (Sco X-1, $\Gamma < -3$) &\\
 tXRBs  & 752   &Spectral ($\Gamma < -2.5$)  & Galactic; $\overline{kT} = 3.2 \pm0.3$  keV\\           
 uGRBs & 65 & Spectral ($\Gamma > -2.5$); Isotropic & Hard; Extragalactic \\
     \tableline
\end{tabular}
\label{tbl:totals}
\end{table*}
\subsection{Thermonuclear X-ray Bursts (tXRBs)}
\label{sec:tXRB}

\begin{figure}[h!]
\centering
    \includegraphics[width=4in]{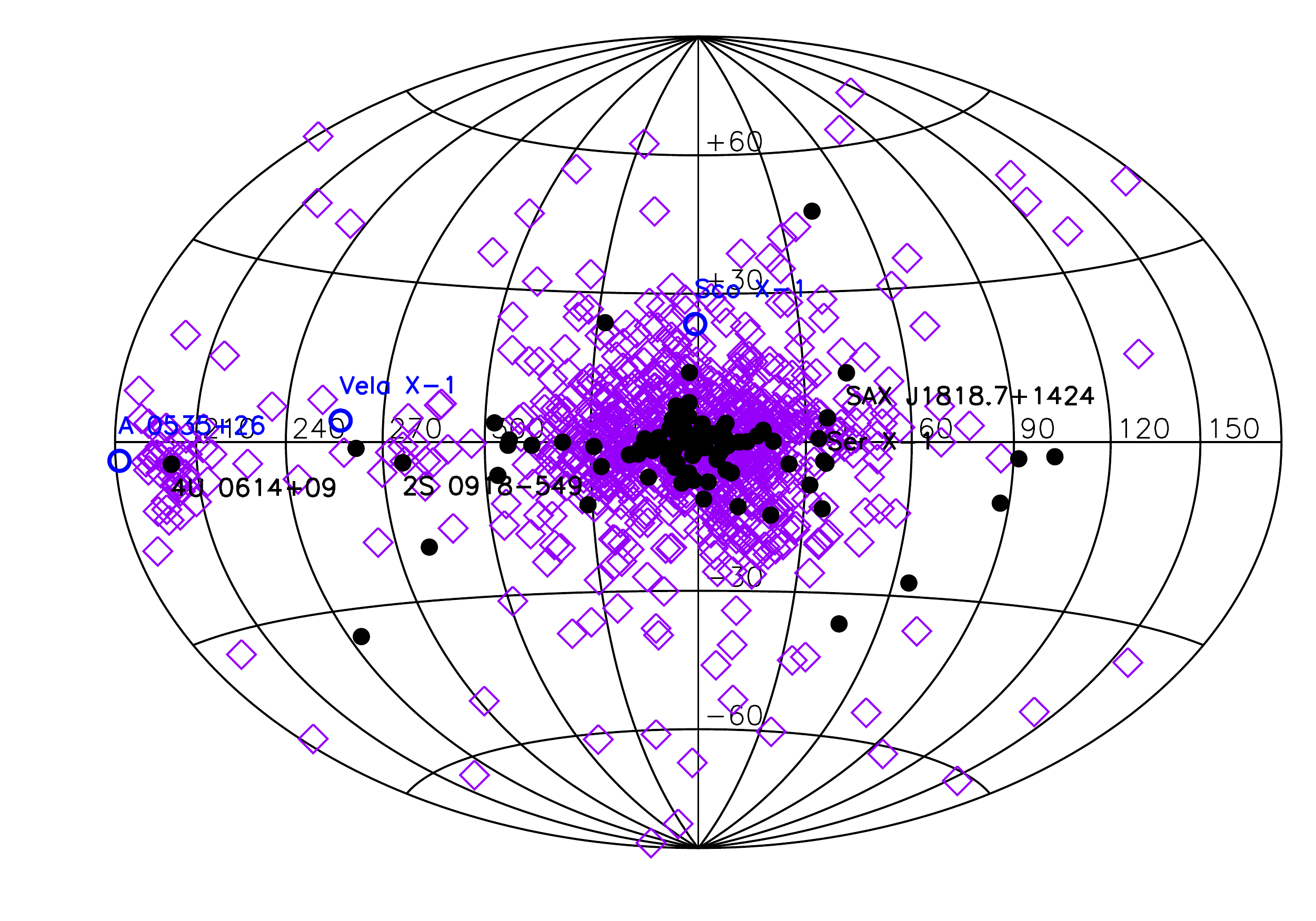} 
    \caption{The purple diamonds are the locations of the tXRBs.  The
      black filled circles are the locations of the known type I
      bursters.  Those that we have multiple associations for are
      labeled in black.  A few of the aFXP sources are
      labeled in blue for comparison.}
  \label{fig:tXRB}
 \end{figure}

The largest category of events in our catalog are soft and their
spectra are well fit using a simple blackbody model with temperature
in the $\sim$2--5~keV range, largely consistent with the spectral
properties of thermonuclear bursts from accreting neutron stars
\citep[e.g.,][]{Swank77}.
They also show a spatial distribution consistent with the $\sim$100
known thermonuclear burst sources (``bursters''; see Figure
\ref{fig:tXRB}), strongly concentrated towards the Galactic bulge
region.
For the bursts that are bright enough, time resolved spectroscopy
reveals cooling along the tail of the burst, the unequivocal signature
of tXRBs. All of our tXRBs associated with 4U 0614+09 are bright
enough for verification via time resolved spectroscopy, including
those reported by \citet{Linares12b}.
In the most energetic bursts from 2S~0918-549 we also detect cooling
along the decay (Section~\ref{sec:2s0918}).

We detect in total 752 tXRB candidates with
375 bright enough for time resolved spectral analysis. Their
average blackbody temperature is 3.2 +/- 0.3 keV.
This value is consistent with the highest temperature measured during
photospheric radius expansion (PRE) bursts, when the photosphere is
thought to reach the neutron star surface at the end of the
Eddington-limited phase \citep[the so-called
``touch-down'';][]{Lewin93,intZand05,Kuulkers_2010}.
The properties of the full tXRBs sample are presented in table
\ref{tbl:type1_table}, including morphology and spectral parameters.
Light curves for these events are given in Appendix \ref{lc:tXRBs},
labeled by burst ID.

GBM location errors are typically larger than a few degrees and
occasionally tens of degrees \citep{Connaughton_2015}.
Since the majority of known bursters are within $\sim$20 degrees
of the Galactic center, individual identification of tXRBs is limited
to those located sufficiently far from the Galactic bulge.
Due to this limitation intrinsic to the GBM location accuracy, we make
no attempt to associate events within this central distribution.
Instead, we focus on those sources which are more than 30 degrees from
Sag A$^*$.
Out of 103 known bursters, this leaves 26 systems that we attempt
to associate with our tXRB events.
Furthermore, we use MAXI 2-20 keV weekly light curves
\citep{Maxi_2009} in an attempt to determine if a given burster was
active at the time of the tXRB (see below).
We place the 26 bursters far from the Galactic bulge into one of the
following four categories.

\begin{itemize}

\item If the source is close enough to be detected in MAXI but has not
  flared within our catalog time period, the source is considered
  always off and we remove it from consideration. Only Cen X-4 is in
  this category.

\item There are 6 bursters that are below the $10\sigma$ detection
  threshold in MAXI but have always shown persistent emission whenever
  they have been observed with pointed X-ray detectors. All but one
  (4U 1323-62, with an orbital period of 2.9~hr) are confirmed or
  candidate ultra-compact X-ray binaries \citep[UCXBs: orbital periods
  shorter than 1~hr; see][]{intZand07}: 4U 0513-40 (in the globular
  cluster NGC~1851), 4U 1246-58, 4U 1915-05 (dipper), 4U 2129+12
  (M15-X2 in the globular cluster M15) and 2S 0918-549 (discussed in
  detail in Sec.~\ref{sec:2s0918}). They are assumed to be
  persistently accreting at a low rate, and considered a candidate for
  association with all events. Their mass accretion rates are
  below 5\% of the Eddington limit \citep{intZand07}, which explains
  the low persistent flux detected by MAXI together with their
  distances $\gtrsim$5~kpc.

\item There are 8 sources whose transient or persistent activity can
  be monitored with MAXI: when actively accreting they are detected
  above the $10\sigma$ threshold. We only consider these sources as
  possible associations to our events if the source exceeds $10\sigma$
  threshold on the week of the event. These sources are 4U 0614+09
  (persistent atoll and UCXB candidate; discussed in detail in
  Sec.~\ref{sec:4u0614}), EXO 0748-676 (quasi-persistent transient, in
  quiescence since 2008), GS 0836-429 (transient, outburst in July
  2012), 4U 1254-69 (persistent atoll dipper), Cir X-1 (peculiar
  atoll/Z), Ser X-1 (persistent), Aql X-1 (canonical atoll transient
  with typically one or two outbursts per year) and Cyg X-2
  (persistent Z source) \citep[see, e.g.,][and references
  therein]{Galloway_2008}.  EXO~0748-676 has not shown activity in MAXI
  during our search period, thus in practice this burster is treated
  as off.

\item The remaining category contains 11 sources with no available
  MAXI weekly light curves. This category includes some of the
  so-called ``burst-only sources'' \citep{Cornelisse02} as well as
  faint transients in which there is only one known outburst with
  which the source was discovered. \emph{Swift}-BAT daily light curves
  for these sources, when available, do not provide a clear
  distinction between quiescent and active periods. These sources are:
  MAXI J1421-613 (outburst in January 2014, i.e., after catalog), UW
  Crb \citep[peculiar persistently faint ``accretion disk corona''
  source, known since 1990][]{Hakala05}, IGR J17062-6143
  \citep[persistently faint at $<$1\% of the Eddington luminosity
  since its discovery in 2006][and references therein]{Degenaar13b},
  SAX J1818.7+1424, SAX J1324.5-6313 and SAX J2224.9+5421
  \citep{Cornelisse02}, Swift J185003.2-005627 (faint transient active
  in May-June 2011), MXB 1906+00, XB 1940-04, XTE J2123-058 and 4U
  2129+47. They are considered for association with the tXRBs, even
  though their activity and mass accretion rate history are often ill
  constrained.

\end{itemize}

An association list is generated for each tXRB using the following
criteria.
If the event location is within $2\sigma$ of a burster in the
association list above, then that event is associated with the source.
%
If more than three sources are associated with an event, then the
event has a large location error, and all associations are rejected as
spurious.
All associations are listed in the table in ascending order
of distance (in $\sigma$ given in parenthesis) from the source.
If only one source locates within $2\sigma$ of an event then it is
listed in bold type, and we consider this a robust association.
%
%
Out of the total of 752 tXRBs, 685 have no
associations and 29 have non-unique associations. We find
unique associations for 54 tXRBs, with eight known bursters.
For this reduced sample we can assess the origin of the bursts,
and their properties are summarized in Table~\ref{tab:totals}.

\subsection{Accretion Flares and X-ray pulses}
\label{sec:aFXP}
\begin{figure}[h!]
\centering
    \includegraphics[width=4in]{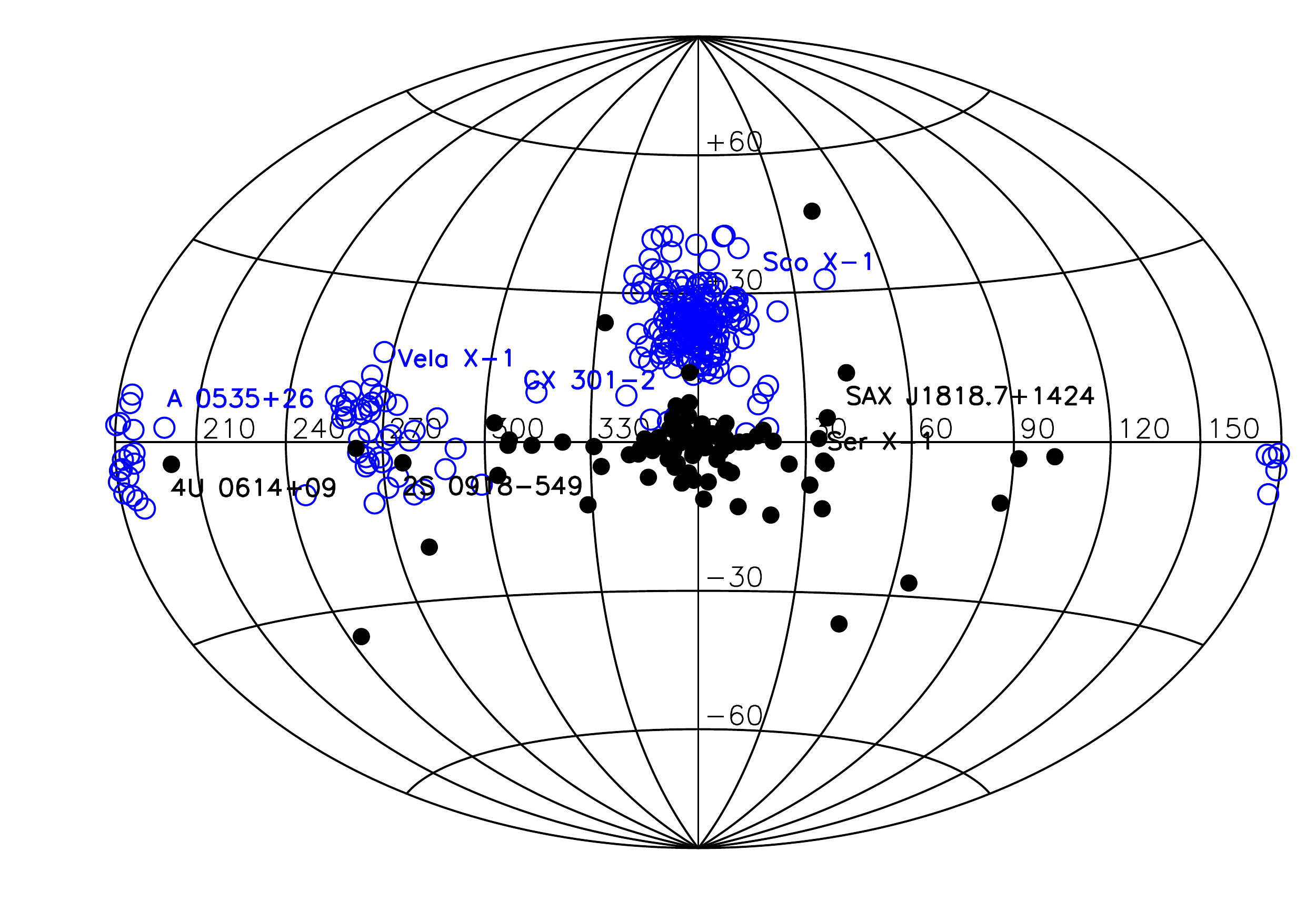} 
   \caption{The blue circles indicate the locations of the aFXPs.  The black filled circles are the locations of the known type I bursters.  Those that we have associations for are labeled in black.}
  \label{fig:aFXP}
 \end{figure}

Accretion powered events such as those originating from Sco X-1, A0535+26 and Vela X-1 are identified once the their location and spectra are know (see Figure \ref{fig:aFXP}).  Sco X-1 events have soft emission (PL index $<$ -3) and are generally well localized to Sco X-1's position.  These events are usually part of a longer flaring episode that is distinctive in GBM channel 1 (12-25 keV) data.  
Events from Vela X-1 and A0535+26 are typically part of a chain of pulsations that are identified in the CTIME data due to the dominate harmonic of their characteristic spin periods of 103.3 s and 283.5 s respectfully as well as their harder spectra with a typical power-law index in excess of -2.5.  The events associated with A 0535+26 coincide with a giant flare from A 0535+26 which occurred in February 2011 \citep{Arranz_2011}. 

The aFXPs are summarized in table \ref{acc_table}.  The columns are as follows: ID is the time of the midpoint, in UTC, of the event selection identified by YYMMDDTTTTT where YY 
indicates the last two digits of the year, MM the month, DD the day, and TTTTT is the time in seconds from the start of the day.  Peak is the time (UTC) of the peak count rate for the event measured in seconds since MJD 55267.  The RA and Dec is the GBM location and the Error is the statistical error on the location. Association is the source which is associated with the event. The light curves for these events are in Appendix \ref{lc:aFXPs} and identified by ID.
\subsection{Untriggered GRBs}
\label{sec:uGRB}
\begin{figure}[h!]
\centering
    \includegraphics[width=4in]{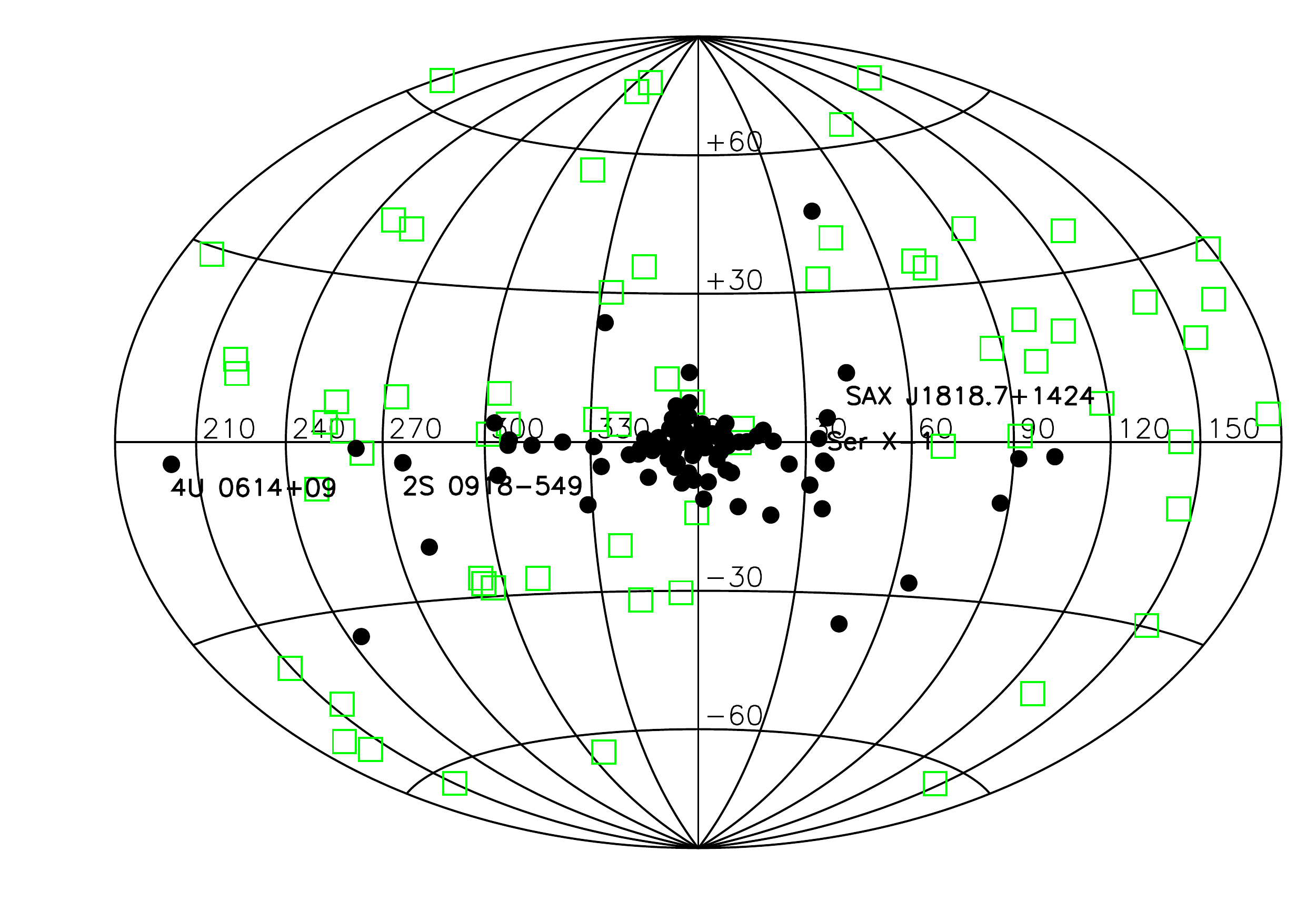} 
   \caption{The green squares indicate the locations of the uGRBs.  The black filled circles are the locations of the known type I bursters.  Those that we have multiple associations for are labeled in black.}
  \label{fig:uGRB}
 \end{figure}

The uGRBs are hard events that are selected due to their isotropic distributed on the sky which implies an extragalactic origin (see Figure \ref{fig:uGRB}).
 In principle extragalactic bursts could arise from sources other than GRBs but given the broad range of spectral and temporal behavior exhibited by GRBs, we use the term uGRB to denote our whole extragalactic population of bursters.      
Their spectra are well fit with a Band function (\cite{Band}) or power-law with an exponential cut-off function that is typical of GRBs.  Parameters from the spectral fits using the Band function were typically not well constrained and are not reported.  The spectral results for the power-law and power-law with exponential cut-off are summarized in Table \ref{untrigg_table}.   The first three columns are the same as the aFXPs.  The next three columns (4-6) are the results of spectral fitting using a power-law with a exponential cut-off parameterized as Epeak.  A `-' in these column denotes that the spectral parameters could not be constrained and these results are left out of the table.  The fifth column is Epeak in keV, the sixth column (Comp Flux) is the energy flux [erg cm$^{-2}$s$^{-1}$] from 10-1000 keV, and the seventh column (Comp Flnc) is the energy fluence [erg cm$^{-2}$] form 10-1000 keV.  The next three columns (8-10) are the results of the spectral fitting using a power-law model.  The eight column is the power-law index, while the ninth and tenth columns are the energy flux and fluence from 10-100 keV.  The last 4 columns are results from the temporal analysis discussed in detail in Section \ref{sec:duration} and include the rise time (Rise), fall time (Fall), duration (Duration), and a column labeled Structure describing the temporal structure of the event.  The Structure column contains an `S' if the light curve is single peaked or an `M' if the light curve is multi-peaked.   If an event is multi-peaked, the rise time and fall time that is calculated may no longer represent a true rise or fall time for the event since the peak of the event could occur on any of the multiple peaks.  The light curves for these events are in Appendix \ref{lc:uGRBs} and identified by ID.

\section{Discussion}
\label{sec:discussion}

We have uncovered a large catalog of untriggered bursts in the GBM data that
reflect the power of GBM as an all-sky monitor of diverse astrophysical phenomena in the hard X-ray energy band. Despite the 
difficulties inherent in uncovering these bursts in the background-limited GBM detectors, and the limitations
imposed by GBM's coarse source localization, we identified at least three distinct classes of events: untriggered GRBs, accretion-powered flares and X-ray pulsations
from known sources, and thermonuclear type I X-ray bursts.

Our source classification relied strongly on spectral modeling and, particularly for the aFXPs, location.
Classification from spectral analysis was complicated by the overlapping distributions of spectral parameters among the different classes.
We used the spatial distribution of source locations on the sky to verify our choice for the spectral hardness cut-off 
for the events assigned to the uGRB sample by verifying that the hardness cut-off maximized the isotropy of the spatial distribution. 

\begin{figure}[h]
 \begin{centering}
    \includegraphics[width=4in]{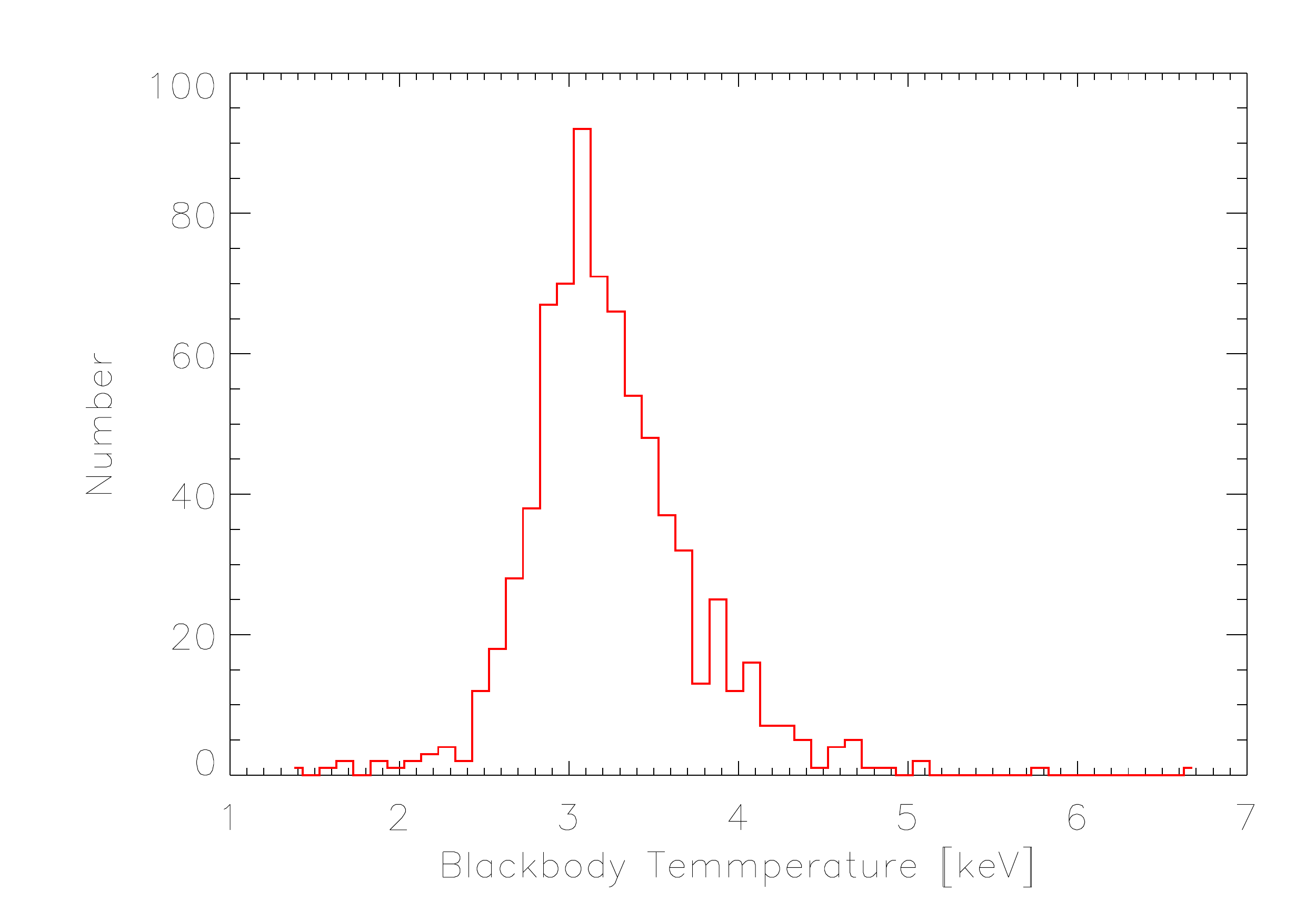}
    \caption{The figure shows the distribution of blackbody temperatures for the tXRBs.  The distribution is not symmetric and has a hard tail that extends to 6 keV.}\label{fig:bb_firsts}
    \end{centering}
\end{figure}

The tXRBs are the primary science driver for this catalog and we discuss them in depth in Section \ref{sec:disc-tXRBs}.  The distribution of temperatures from the blackbody spectral fits of the tXRBs is shown in Figure \ref{fig:bb_firsts}.  The temperature distribution has a hard tail that extends beyond 6 keV prompting speculation that there was a fourth, unknown, category of XRBs.  Monte Carlo analysis performed in Section \ref{sec:spetral} indicates that GBM has sufficient spectral sensitivity to accurately measure the spectral temperature down to 3 keV yet there are 62 tXRBs whose spectral temperature exceed 4.0 keV and none are associated with a known type I source.  Furthermore, these events are distributed along the Galactic plane and concentrated at the Galactic center.  A few are weak and may be explained by poor background subtraction while a few may be soft GRBs with chance location along the Galactic plane.  We find no evidence of a bimodal distribution in spectral temperature or fluence thus we conclude that a fourth `unknown' category is unwarranted with the current data set.  We will revisit this when more data has been analyzed.

The aFXPs were a byproduct of our XRB search since we have dedicated programs to study them (The GBM Pulsar Project\footnote{http://gammaray.msfc.nasa.gov/gbm/science/pulsars.html} and The Earth Occultation Project\footnote{http://heastro.phys.lsu.edu/gbm}).  Nevertheless; the aFXPs in the catalog provide a unique opportunity to observe these sources in rare, bright states that would normally require a targeted observation.

The brightest (other than the Sun) recurring source GBM observes from 8-50 keV is Sco X-1 and we intentionally attempted to avoid this source since our focus was on tXRBs, nevertheless; Sco X-1 dominates the aFXP category due to its numerous flares.  Its persistent nature makes background subtraction difficult and this occasionally leads to poor localization.  Luckily, its soft spectrum (index $\sim -3.5$) makes this source relatively easy to identify.  The other aFXPs are magnetically dominated accretion powered neutron stars with a harder spectrum (index $\sim -2$).  Again, none of the aFXPs were intentionally targeted by our efforts but bright pulsations from these sources occasionally mimic XRBs in the 12-25 keV band and only careful follow-up review of these events reveal the train of pulses that help identify these sources.    

\begin{figure}[h]
 \begin{centering}
    \includegraphics[width=4in]{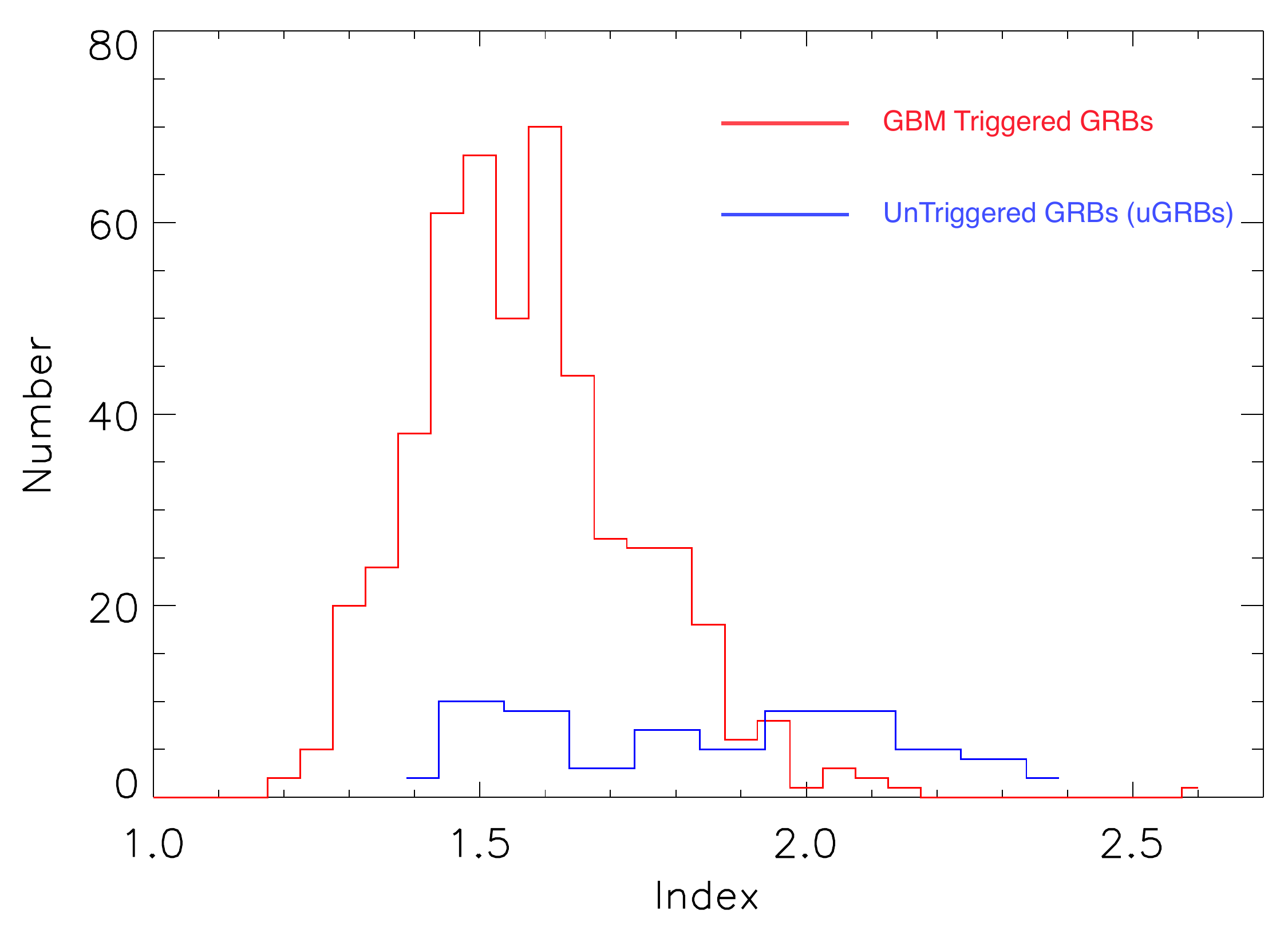}
    \caption{The red curve shows the histogram of the spectral index of comparable GBM triggered long GRBs while the blue curve is the histogram of the spectral index of the uGRBs in the XRB catalog}\label{fig:GRB_comp}
    \end{centering}
\end{figure}

The uGRBs are either GBM sub-threshold trigger events or events which occur when triggering is disabled (rare occurrence).  The sub-threshold events are an interesting population of GRBs which might include intrinsically weak, distant, or off-axis GRBs whose detection has consequences for population synthesis studies and future gravitational wave experiments optimized to detect rotating collapsars \citep{Ott_2011} and will be explored in future
work.  Figure \ref{fig:GRB_comp} shows the spectral index distribution (in red) of the GBM triggered GRBs during the XRB catalog period whose duration (T90) is greater than four seconds.  Overlaid (in blue) on the triggered distribution are the uGRB's spectral index distribution.  It is reasonable from the figure to claim that most of the uGRBs are a sub-threshold continuation of the the triggered GRB population.


\subsection{GBM's view on thermonuclear bursts}
\label{sec:disc-tXRBs}

With an instantaneous FoV covering 75\% of the sky, GBM offers an
unprecedented coverage of most Galactic bursters.
Due to its sensitivity at energies above $\sim$8~keV, GBM detects only
the hottest phases of the hottest type I X-ray bursts: the touch-down
phase of PRE bursts (as shown quantitatively in the simulations
presented in \citet{Linares12b})
Thus our GBM X-ray burst monitor is a ``PRE burst monitor'' with an
excellent observing duty cycle (50\%, only interrupted by Earth
occultations and SAA passages).

Figure \ref{fig:txrbs} shows a histogram of energy flux (10--100 keV)
for the tXRBs from the blackbody spectral fits.
The flux distribution was fit ($\chi^2 = 101/81$ dof) with a gaussian with the centroid at
3.1$\times10^{-8}$ erg cm$^{-2}$ s$^{-1}$ and a standard deviation of 1.2$\times10^{-8}$ erg
cm$^{-2}$ s$^{-1}$.
The faintest tXRB in the catalog has a flux of
(3.4$\pm$1.0)$\times10^{-9}$ erg cm$^{-2}$ s$^{-1}$, which gives an
estimate of the absolute flux limit in our catalog.
Due to the strongly variable X-ray background at 8--50~keV; however,
the minimum detectable flux can vary strongly.

PRE bursts reach the Eddington limit, which for a 1.4~M$_\odot$
neutron star is in the range [1.6--3.8]$\times$10$^{38}$~erg~s$^{-1}$
\citep[depending of the radius and composition of the photosphere;
see, e.g.,][]{Lewin93,Kuulkers_2003}.
In order to test if the flux distribution is consistent with
thermonuclear bursts from the Galactic bulge, we adopt a fiducial
Eddington luminosity of L$_{Edd}$=2.5$\times$10$^{38}$~erg~s$^{-1}$,
and show in Figure~\ref{fig:txrbs} the 10--100~keV fluxes
corresponding to L$_{Edd}$ at a distance of 8~kpc and 10~kpc
(horizontal lines labelled L$_{Edd,8}$ and L$_{Edd,10}$ and).
At least three factors contribute to the observed flux scatter: i)
bursters have a range of distances, ii) different systems can have
different L$_{Edd}$ (due to differences in neutron star mass, radius
or photospheric composition), and iii) even in a given burster the
peak luminosity of PRE bursts show significant scatter
\citep{Galloway_2008}.

We thus conclude, from the flux distribution shown in
Figure~\ref{fig:txrbs}, that our tXRB sample is consistent with a
population of Eddington-limited PRE bursts coming from a mix of
bursters around the Galactic bulge region.
Moreover, because only a handful of tXRBs have fluxes lower than that
corresponding to an Eddington-limited burst at 10~kpc (yet several
known bursters are farther than that), we estimate that our catalog is
limited to PRE bursts occurring within $\sim$10~kpc.
The fluence distribution, on the other hand, shows that for an assumed
distance of 8~kpc, most bursts have energies between 10$^{39}$~erg and
10$^{40}$~erg (see horizontal lines labelled E39 and E40 in
Fig.~\ref{fig:txrbs}), although the range of fluences is wide with
about two orders of magnitude.
The duration of the tXRBs in the GBM band also spans a wide range,
between $\sim$5~s and $\sim$500~s.
The observed distributions of fluence and duration are not bimodal,
both in the full tXRB sample and in the two low-$\dot{m}$ bursters
presented in Sections~\ref{sec:4u0614} and \ref{sec:2s0918}.
This indicates that the longest and most energetic thermonuclear
bursts, sometimes referred to as ``intermediate/long bursts'', are an
extreme case of normal burst ignition.

We use hereafter a bolometric correction factor of f$_{bolo}$=1.9 to
convert from 10--100~keV to bolometric burst flux and fluence, which
we derive using a typical kT$_{bb}$=3~keV spectrum.
Moreover, due to the high background rate and lack of sensitivity
below 8~keV, GBM only detects the peak of tXRBs, where the temperature
is highest.
To take this into account (i.e., to include an estimate of the energy
radiated during the burst tail), we use a ``band correction factor''
of f$_{band}$=1.3 to convert from GBM fluences (8--50~keV) to a more
standard (2--50~keV) energy band.
This band correction was calculated by \citet{Linares12b} using
simulated GBM lightcurves of bursts observed with the RXTE-PCA.

\begin{table*}[h!]
\footnotesize
\caption{GBM bursts with associated bursters}
\begin{minipage}{\textwidth}
\begin{tabular}{l c c c c c c c}
\tableline\tableline
 Burster & Nr.Bursts & D(kpc) &  L(10$^{38}$~erg/s) & E(10$^{39}$~erg) & dur(s) & rise(s) & $<$kT$>$(keV) \\
\tableline

4U 0614+09 & 33 & 3.2\footnote{\citep{Kuulkers_2010} } & 	0.3-1.7 & 0.4-6.1 & 6.0-51.3 & 1.2-8.6 & 3.2\\
2S 0918-549 & 10 & 5.0\footnote{\citep{intZand05} 4.0-5.3 kpc} & 	0.7-1.7 & 1.0-17.0 & 9.7-75.6 & 2.5-38.9 & 3.1\\
SAX J1818.7+1424 & 4 & 9.4\footnote{\citep{Cornelisse02} $<$9.4 kpc} & 	2.2-4.3 & 8.8-25.7 & 20.1-90.9 & 13.5-62.0 & 3.5\\
UW Crb & 2 & 5\footnote{\citep{Hakala05} $>$5-7 kpc} & 	1.0-1.0 & 1.4-1.8 & 10.6-13.7 & 2.4-4.9 & 3.2\\
IGR J17062-6143 & 2 & 5\footnote{\citep{Degenaar13b} } & 	0.9-1.0 & 2.0-2.6 & 16.0-22.0 & 6.9-10.2 & 3.6\\
XB 1940-04 & 1 & 8\footnote{\citep{Murakami83} unknown distance} & 	4.4-4.4 & 28.9-28.9 & 50.1-50.1 & 11.8-11.8 & 3.9\\
Ser X-1 & 1 & 8.4\footnote{\citep{Cornelisse02b} 7.7-10.0 kpc} & 	2.8-2.8 & 19.4-19.4 & 52.6-52.6 & 39.9-39.9 & 3.1\\
MAXI J1421-613 & 1 & 7\footnote{\citep{Serino15} $<$7 kpc} & 	2.0-2.0 & 13.8-13.8 & 53.9-53.9 & 2.3-2.3 & 4.8\\
\tableline\tableline
\end{tabular}
\end{minipage}
\label{tab:totals}
\end{table*}

\subsubsection{4U 0614+09}
\label{sec:4u0614}

The burster and UCXB candidate 4U~0614+09 has been
extensively studied by most X-ray missions, and is known to accrete
persistently at a rate close to 1\% of the Eddington limit.
Due to its location far from other bursters and its proximity
(\citealt{Kuulkers_2010} measured a distance of d=3.2~kpc, which we adopt
in this work), it is an ideal source to study thermonuclear
bursts at low accretion rates. During the first year of the
Fermi-GBM X-ray burst monitor we detected 15 bursts from 4U~0614+09
\citep{Linares12b}.

\begin{figure}[h!]
\centering
    \rotatebox{-90}{\includegraphics[width=4in]{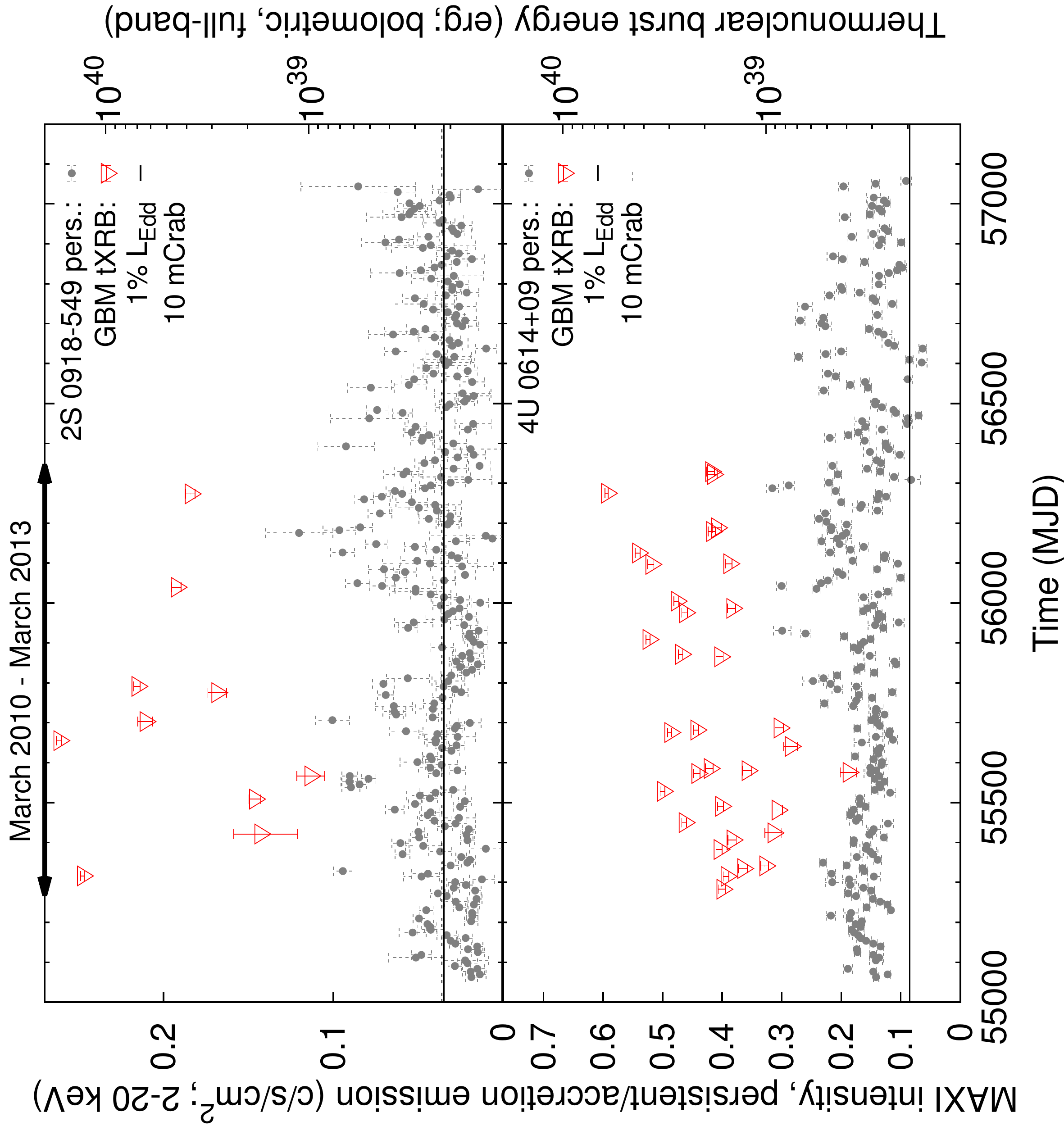} }
    \caption{Red triangles show the time and radiated energy (right
      axis scale, band-corrected; see text) of the tXRBs from
      4U~0614+09 {\it(bottom)} and 2S~0918-549 {\it (top)} detected by
      GBM during the period covered by this catalog (shown with the
      black arrow). The X-ray intensity is shown in the same panels
      (gray small circles, left axis), tracing the mass accretion rate
      history of each burster. The solid and dashed horizontal lines
      show the corresponding 1\% L$_{Edd}$ and 10 mCrab levels,
      respectively.}
\label{fig:lcs}
\end{figure}

Our three year catalog includes 33 tXRBs from 4U 0614+09 detected by
GBM between March 2010 and March 2013. This is the same number of
bursts detected from 4U~0614+09 with 9 different instruments over the
course of 15 years \citep[1992--2007,][]{Kuulkers_2010}, which shows the
drastic improvement in detection efficiency gained by GBM.
Given GBM's 50\% observing duty cycle, we measure a burst recurrence
time of t$_{rec}$=17$\pm$2~d (1$\sigma$ Poissonian uncertainty), 5~d
longer than, but consistent with, the results of \citet{Linares12b}.
The closest burst pair we find is only 1.4~d apart, on 2012-06-18/20,
the shortest wait time between thermonuclear bursts measured from this
source to date.
The bolometric and band-corrected burst energies from 4U~0614+09 span
more than an order of magnitude, between
[0.4--6.1]$\times$10$^{39}$~erg, and show no evidence of bimodality,
as shown in Table~\ref{tab:totals} and Figure~\ref{fig:lcs}.

\subsubsection{2S 0918-549}
\label{sec:2s0918}

Before our GBM campaign, 7 thermonuclear bursts had been reported from
the UCXB candidate 2S~0918-549, between 1996 and 2004 \citep[][we use
the same distance of d=5~kpc throughout this work]{intZand05}.
This burster is analogous to 4U~0614+09 in many ways: both are
candidate UCXBs persistently accreting at a very low rate, and without
detected hydrogen or helium lines in the optical spectrum
\citep{Nelemans04}.
The inferred mass accretion rate in 2S~0918-549 is about two times
lower, $\sim$0.5\% of the Eddington limit (see Figure~\ref{fig:lcs}).
In three years, we detect 10 tXRBs from
2S~0918-549, yielding a recurrence time of t$_{rec}$=56$\pm$12~d.
The closest pair of bursts was detected in August 2011, only
$\sim$16~d apart.

 \begin{figure}[h!]
\centering
    \includegraphics[width=4in]{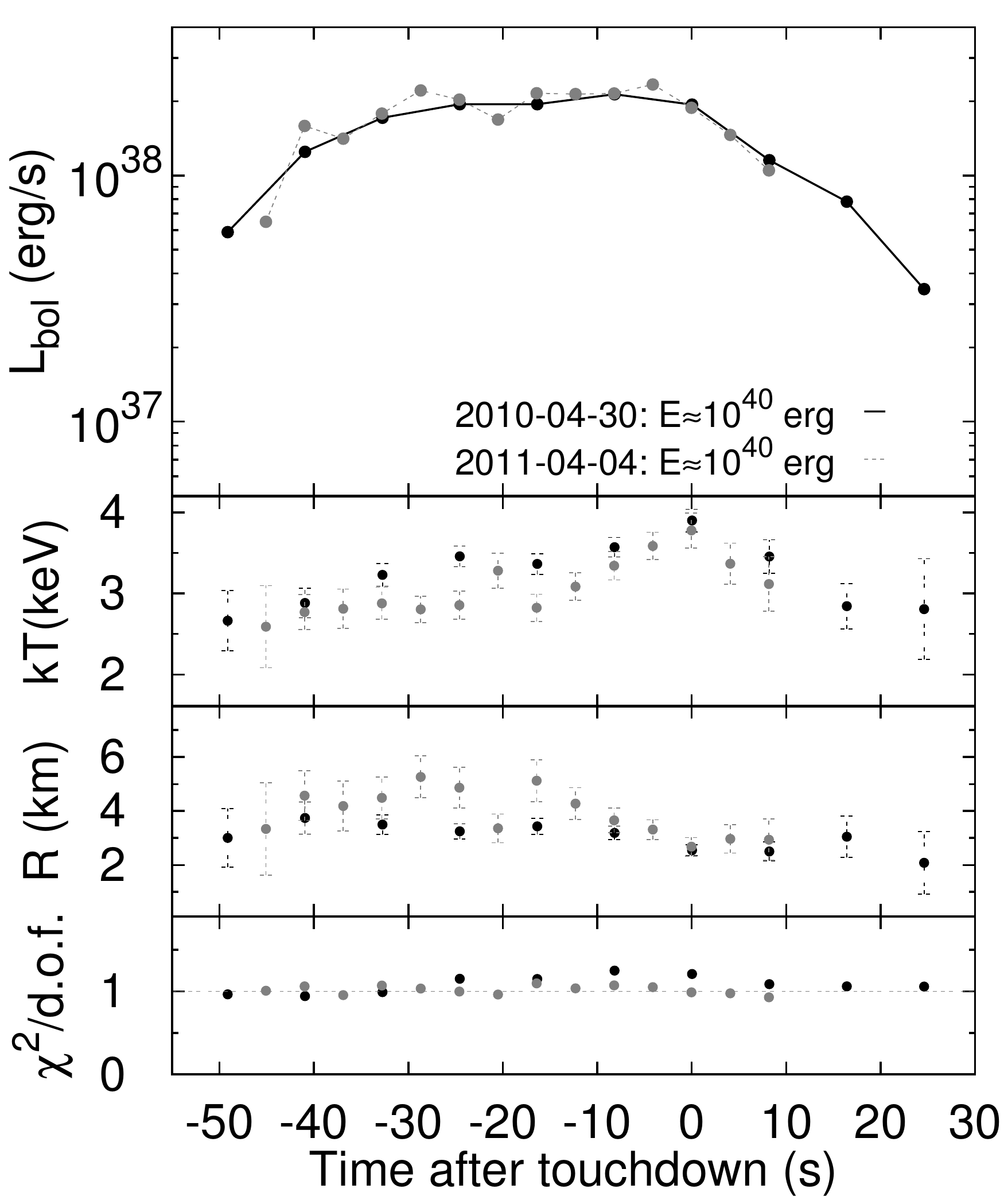}
    \caption{Luminosity and spectral evolution of the two most
      energetic bursts from 2S~0918-549, from top to bottom panels:
      bolometric luminosity, blackbody temperature, apparent emmiting
      radius and reduced chi squared. The maximum temperature and
      highest signal-to-noise in the GBM detectors coincides with the
      so-called ``touch-down'', when the neutron star photosphere is
      thought to reach the surface after expanding and contracting.}
\label{fig:2s0918}
\end{figure}

Two of the bursts, shown in Figure~\ref{fig:2s0918} are consistent
with the so-called ``long bursts''
\citep{intZand05,Cumming06,Chenevez08}. These were detected on
2010-04-30 and 2011-04-04, with durations in the GBM band of 66~s and
76~s, respectively.
They have energies above 10$^{40}$~erg and their duration in the full
2--50~keV band is likely more than ten times longer than that in the
GBM band, i.e., several tens of minutes \citep[the band-corrected
duration is more uncertain than the total energy, see][for an
estimate]{Linares12b}.
Both the burst durations and energies show a continuous distribution
in the range [10--76]~s and [0.1--1.7]$\times$10$^{40}$~erg,
respectively (energies are bolometric and band-corrected, see
Sec.\ref{sec:tXRB}).

\subsubsection{Other bursters and the integrated Galactic tXRB sample}
\label{sec:others}

The remaining associations, 11 tXRBs detected from the direction of
six other bursters, are presented in Table~\ref{tab:totals}.
Some of these events are faint and have large location errors
(Table~\ref{tbl:type1_table}), which together with the low number of
tXRBs per burster makes the association uncertain.
These include: i) four events from the direction of SAX J1818.7+1424
(detected on 2010-07-01, 2010-07-11, 2010-10-02 and 2011-01-28);
ii) two events from the direction of the high Galactic latitude
burster UW Crb (on 2011-11-03 and 2011-12-31) and two from the
direction of IGR J17062-6143 (on 2010-07-19 and 2011-04-29);
iii) one event associated to XB 1940-04 (2011-10-20), one to Ser X-1
(2010-05-31), and one tXRB from the direction of MAXI J1421-613 (on
2011-10-16; note that this source was discovered in outburst in
January 2014).

It is also worth discussing which bursters are {\it missing} from the
association list.
Most notoriously, we do not detect any tXRB from 4U~1246--58 in our
three-year catalog.
In a study of this burster and UCXB candidate accreting persistently
below 1\% of the Eddington rate, \citet{intZand08} found 7 PRE bursts,
all but two with long durations, and a distance of 4.3~kpc.
The corresponding burst rate between 1996 and 2008 was 12$\pm$6 d
\citep{intZand07}.
In contrast, our non-detection of GBM bursts from 4U~1246--58 between
2010 and 2013 implies a 95\% lower limit on the recurrence time
t$_{rec}$$>$186~d, significantly longer than that measured by
\citet{intZand07}.
Thus our results suggest that a drastic change in the burst properties
of this burster took place between 2008 and 2010, which might be
linked to the long-term decay of its persistent emission already noted
in \citet{intZand08}.



Two other UCXB bursters are probably too distant to be detected with the
GBM X-ray burst monitor: 4U 0513-40 \citep[8.2--11~kpc according
to][]{Galloway_2008} and 4U~2129+12 (X-2 in M15, 10.4 kpc away according
to \citealt{Harris96}).
4U~1915-05 is also close to our detection limit \citep[6.8--8.9 kpc
according to][]{Galloway_2008} and is a high inclination ``dipper'' UCXB,
which may explain why no bursts are detected by GBM in the present
catalog.
The rest of UCXBs and low mass accretion rate bursters are too close
to the extended Galactic bulge region to be resolved by GBM, but are 
included in the total Galactic rate measured and discussed
below.

 \begin{figure}[]
\centering
    \rotatebox{-90}{\includegraphics[width=4in]{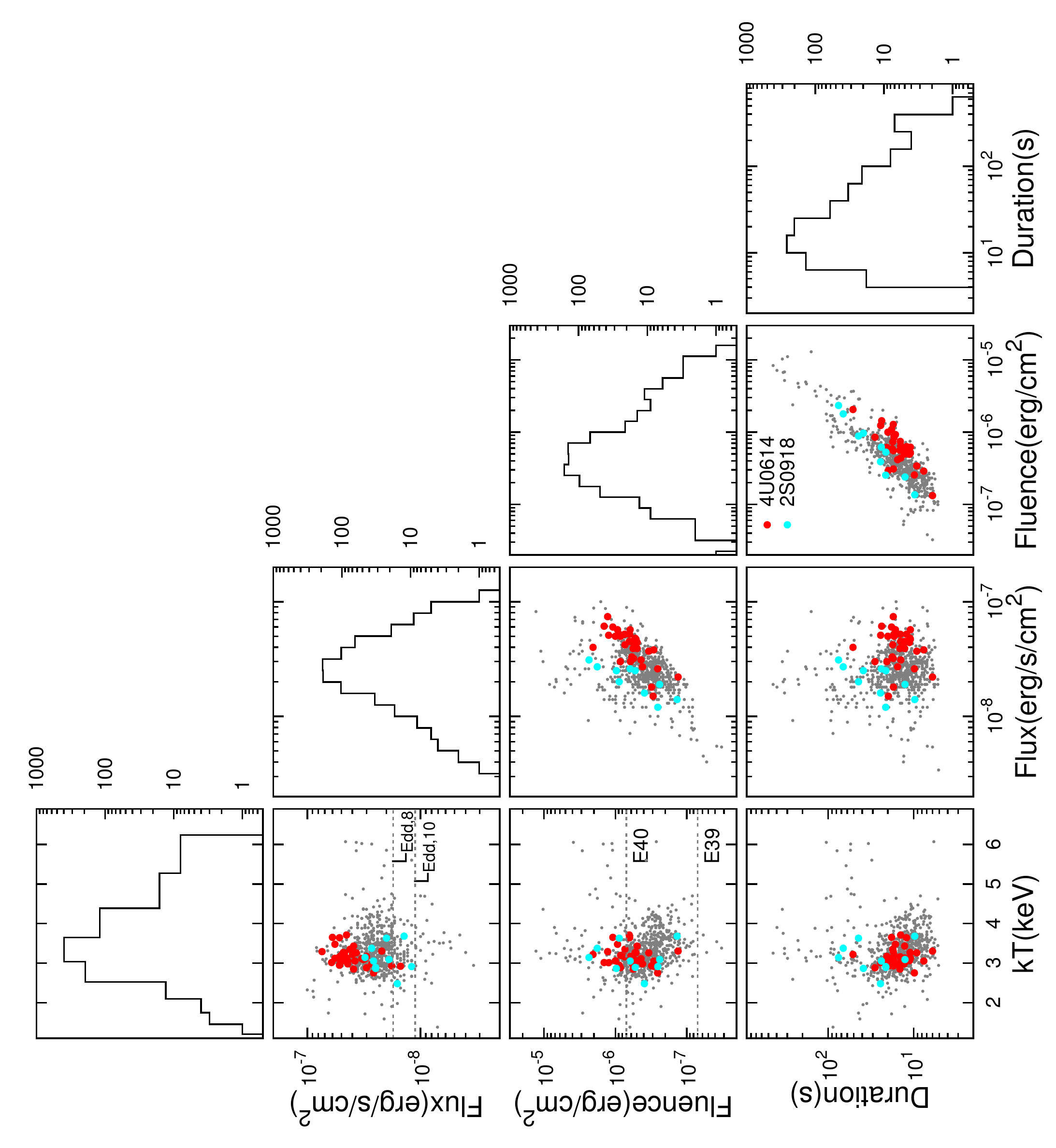} }
    \caption{Temperature, flux (10--100~keV), fluence (10--100~keV)
      and duration distribution of the full tXRB sample, as measured
      by GBM. The bursts from 4U~0614+09 and 2S~0918-549 are shown in
      red and blue, respectively. Horizontal dashed lines show, for
      comparison and from top to bottom: GBM flux corresponding to an
      Eddington luminosity L$_{Edd}$=2.5$\times$10$^{38}$~erg~s$^{-1}$
      at a distance of 8~kpc; GBM fluence corresponding to a
      bolometric energy of 10$^{40}$ and 10$^{39}$~erg at
      8kpc. Histograms of all four parameters are also shown.}
\label{fig:txrbs}
\end{figure}

We show in Figure~\ref{fig:txrbs} the distribution of blackbody
temperature, flux, fluence and duration, in the full sample of tXRBs.
The vast majority of tXRBs come from the Galactic ``extended bulge''
region (489 locate to within 30$^\circ$ of Sag-A).
Fluence and duration are clearly correlated, showing that the most
energetic thermonuclear bursts are also the longest, as expected given
the physical (Eddington) limit on the burst luminosity.

The total of 752 tXRBs detected in our three-year catalog, correcting
for the 50\% observing duty cycle, implies a total rate of
1.37$\pm$0.04 thermonuclear bursts per day (1-$\sigma$ Poissonian
uncertainty).
This represents the average over three years of all bursters within
the reach of the GBM X-ray burst monitor, which we estimate below
corresponds to distances $\lesssim$10~kpc
(Section~\ref{sec:discussion}).
Due to GBM's broad sky coverage, this constitutes an
unprecedented measurement of the total Galactic thermonuclear burst
rate, which we discuss in Section~\ref{sec:discussion}.

On average, the GBM bursts in 2S~0918-549 (t$_{rec}$=56$\pm$12~d;
$<$E$>$=6$\times$10$^{39}$~erg) are more energetic and less frequent
than those from 4U~0614+09 (t$_{rec}$=17$\pm$2~d;
$<$E$>$=2$\times$10$^{39}$~erg).
This is qualitatively explained by ignition models, given that
2S~0918-549 accretes at a rate about two times lower than 4U~0614+09
\citep{Cumming06}.
Lower $\dot{m}$ implies a colder neutron star envelope, a longer fuel
accumulation time and a higher ignition depth.
However, ignition models still have problems to reproduce
quantitatively these recurrence times and burst energies
\citep{Kuulkers_2010,Linares12b}.
Assuming Solar metallicity, the pure helium ignition models from
\citet{CB00} require large amounts of deep crustal heating to
reproduce the recurrence times that we measure in 2S~0918-549 and
4U~0614+09: more than 3~MeV per accreted nucleon \citep[see Figure~7
and further discussion in][]{Linares12b}.
Having two low-$\dot{m}$ bursters with robust GBM measurements of
recurrence times, we can place the first meaningful constraints on the
t$_{rec}$--$\dot{m}$ relation at $\dot{m}/\dot{m}_{Edd} \sim 1\%$.
The measured t$_{rec}$ and $\dot{m}$ in 2S~0918-549 and 4U~0614+09 are
not consistent with a linear relation, and suggest a steeper
t$_{rec} \propto \dot{m}^{[1.7-1.8]}$ relation.

We find a total Galactic rate of 1.4 PRE bursts per day, out to about
10~kpc from the Sun and averaged over the three years of our catalog
(Section~\ref{sec:others}).
During PRE bursts the neutron star atmosphere can be pushed by
radiation forces up to hundreds or thousands of kilometers above the
surface, and small but significant amounts of nuclear burning ashes
may be ejected \citep{Weinberg06}.
To conclude, we roughly estimate the total mass ejected by the
PRE bursts uncovered by GBM, by adding their bolometric- and
band-corrected fluences and assuming they are all at 8~kpc (see
Sec.~\ref{sec:tXRB} and discussion above).
This yields a total radiated energy of 8$\times$10$^{42}$~erg,
1.6$\times$10$^{43}$~erg after correcting for the 50\% observing duty
cycle.
For a nuclear energy release of 1.6--4.4 MeV~per nucleon, this
translates into [4--11]$\times$10$^{24}$~gr of burned fuel.
For a fraction of ejected mass of 10$^{-4}$--10$^{-2}$
\citet{Weinberg06}, this implies a total of
[4$\times$10$^{20}$--10$^{23}$~gr] ejected during three years.
With the above assumptions, we are able to place direct observational
constraints on the amount of mass ejected into the interstellar medium
by PRE bursts in our Galaxy (within 10~kpc of the Sun):
10$^{-13}$--10$^{-11}$~M$_\odot$~yr$^{-1}$.
Whether or not this contributes significantly to the Galactic
abundances of any elements \citep[proton-rich isotopes have received
  particular attention in the context of thermonuclear bursts;
  see][and references therein]{Weinberg06}, remains a subject for
future studies.

\acknowledgments{\bf{Acknowledgements}}
\begin{itemize}
\item M.L. was supported by the Spanish Ministry of Economy and Competitiveness under the grant AYA2013-42627.\\
\item This work was also supported by NASA Fermi-GI grant nr. NNX11AO19G (PI: Linares). \\
\item This research has made use of the MAXI data provided by RIKEN, JAXA and the MAXI team.\\
\end{itemize}

\clearpage
\pagenumbering{gobble}
\begin{tiny} 
 \begin{center} 
\begin{longtable}{lccccc}   
\caption[Sources]{GBM Accretion Powered Events} \label{acc_table}
 \\
 \hline 
 \label{sources}
 \centering
 \label{tbl:main}
ID & Peak & Ra & Dec & Error & Association   \\ 
& s &degrees &degrees &degrees &   \\
\hline
\endfirsthead 
 \caption{GBM Accretion Powered Events (continued from previous page)}  \\
 \hline  
ID & Peak & Ra & Dec & Error & Association  \\ 
& s& degrees &degrees &degrees &   \\
\hline
  \endhead
10033115145 & 01656753 & 237.6 & -19.0 & 6.6 & ScoX-1\\
10033122117 & 01663724 & 253.6 & -10.7 & 6.4 & Sco X-1\\
10041602529 & 03026531 & 239.0 & -19.0 & 1.1 & ScoX-1\\
10041602579 & 03026532 & 240.2 & -12.1 & 1.6 & ScoX-1\\
10041602801 & 03026803 & 250.2 & -21.8 & 2.1 & ScoX-1\\
10041610060 & 03034062 & 245.7 & -13.8 & 2.1 & Sco X-1\\
10041610234 & 03034236 & 245.7 & -18.5 & 6.1 & Sco X-1\\
10042668136 & 03956138 & 246.9 & -12.1 & 4.0 & ScoX-1\\
\end{longtable}
\end{center}
\end{tiny}

 \setlength{\tabcolsep}{4pt}   \begin{landscape}
 \begin{tiny} 
 
\begin{center}

  \begin{longtable}{lccccccccccccccc}  
\caption[Sources]{GBM Untriggered GRB Events} \label{untrigg_table}\\
\hline   \label{sources} \centering  \label{tbl:main}
ID &Peak  & Ra & Dec & Error & Epeak &Comp Flux & Comp Flnc & PL Index & PL Flux & PL Flnc & Rise & Fall  & Duration & Structure\\
\ \ \
 & s &&   & (sigma) & keV& $10{-8}$ erg cm$^{-2}$  $s^{-1}$&  $10^{-7} $erg cm$^{-2}$ & & $10^{-8}$erg cm$^{-2}$ s$^{-1}$ & $10^{-7}$ erg cm$^{-2}$ & sec & sec & sec & \\
 \hline    \endfirsthead   \caption{GBM Untriggered GRB Events (continued from previous page)}  \\  \hline  
 ID & Peak &Ra & Dec & Error & Epeak &Comp Flux & Comp Flnc & PL Index & PL Flux & PL Flnc & Rise & Fall & Duration & Structure\\
 \ \ \
 &s& &   &(sigma) & keV& $10{-8}$ erg cm$^{-2}$  $s^{-1}$&  $10^{-7} $erg cm$^{-2}$ & & $10^{-8}$erg cm$^{-2}$ s$^{-1}$ & $10^{-7}$ erg cm$^{-2}$ & sec & sec & sec & \\
 \hline   \endhead
10031206566 & 00006572 & 94.6 & 71.7 & 6.7 & $48148.250000 \pm 7221517.$ & 
$7.17 \pm 0.90$ & $20.5 \pm 2.5$ & $-1.675 \pm 0.064$ & $7.82 \pm 0.85$ & 
$22.3 \pm 2.4$ &   9.38 &  10.67 &  21.34 & S\\
10040364545 & 01965350 & 67.2 & -13.4 & 9.8 & $63.8 \pm 10.$ & $5.09 \pm 0.56$
 & $14.5 \pm 1.6$ & $-1.577 \pm 0.099$ & $13.3 \pm 2.6$ & $38.2 \pm 7.5$ & 
  3.48 &  21.87 &  27.25 & S\\
10041528253 & 02965835 & 261.6 & 47.8 & 2.1 & $24.2 \pm 2.0$ & $2.28 \pm 0.23$
 & $17.7 \pm 1.8$ & $-2.17 \pm 0.15$ & $4.28 \pm 0.96$ & $33.2 \pm 7.4$ &  28.38
 &  95.49 & 125.40 & S\\
10041565751 & 03003352 & 272.7 & -19.1 & 4.9 & $92.0 \pm 12.$ & $7.99 \pm 0.72$
 & $22.8 \pm 2.0$ & $-1.595 \pm 0.057$ & $15.3 \pm 1.4$ & $43.8 \pm 4.2$ & 
  7.65 &   9.79 &  19.38 & S\\
10062847601 & 09378789 & 358.8 & 69.1 & 1.0 & $125.6 \pm 7.0$ & $15.91 \pm 0.62$
 & $136.3 \pm 5.3$ & $-1.492 \pm 0.020$ & $28.55 \pm 0.96$ & $244.6 \pm 8.2$ & 
 24.75 &  63.99 & 101.82 & S\\
10071023190 & 10391193 & 299.8 & -45.0 & 12.5 & $69.9 \pm 8.5$ & $5.14 \pm 0.47$
 & $10.50 \pm 0.97$ & $-1.725 \pm 0.070$ & $9.4 \pm 1.0$ & $19.3 \pm 2.1$ & 
  9.29 &   5.32 &  15.33 & M\\
10071069939 & 10437939 & 89.9 & -84.0 & 5.4 & $62.6 \pm 11.$ & $3.94 \pm 0.44$
 & $9.6 \pm 1.0$ & $-1.799 \pm 0.075$ & $6.94 \pm 0.89$ & $16.9 \pm 2.1$ & 
  5.52 &   8.74 &  14.66 & M\\
10071369978 & 10697195 & 290.2 & 68.3 & 15.9 & $46.5 \pm 15.$ & $3.46 \pm 0.63$
 & $7.0 \pm 1.3$ & $-1.99 \pm 0.12$ & $5.18 \pm 0.99$ & $10.5 \pm 2.0$ &  14.30
 &   7.95 &  23.36 & M\\
\hline
\end{longtable}
\end{center}
\end{tiny}
\end{landscape}

  \setlength{\tabcolsep}{3pt}   
\begin{landscape}
\begin{tiny} 
\begin{center}
 \begin{longtable}{lccccccccccccccc}  
\caption[Sources]{GBM Type 1 Events}  \label{tbl:type1_table} \\
\hline   \label{sources} \centering  \label{tbl:main} 
ID&Peak & Ra & Dec & Error & Name (distance) & BB temp &BB flux & BB Flnc & PL index & PL Flux & PL Flnc & Rise & Fall  & Duration & Structure\\
\ \ \
 &s& & &  & (sigma) & keV& $10{-8}$ erg cm$^{-2}$  $s^{-1}$&  $10^{-7} $erg cm$^{-2}$ & & $10^{-8}$erg cm$^{-2}$ s$^{-1}$ & $10^{-7}$ erg cm$^{-2}$ & sec & sec & sec & \\
 \hline    \endfirsthead   \caption{GBM Type 1 Events continued from previous page}  \\  \hline  
 ID&Peak & Ra & Dec & Error & Name(distance) & BB temp &BB flux & BB Flnc & PL index & PL Flux & PL Flnc & Rise & Fall & Duration & Structure\\
 \ \ \
 & s & & &  &(sigma) & keV& $10{-8}$ erg cm$^{-2}$  $s^{-1}$&  $10^{-7} $erg cm$^{-2}$ & & $10^{-8}$erg cm$^{-2}$ s$^{-1}$ & $10^{-7}$ erg cm$^{-2}$ & sec & sec & sec & \\
 \hline   \endhead
10032800979 & 01383378 & 98.4 & 4.6 & 3.4 & \bf{4U\_0614+09} & $3.09 \pm 0.10$
 & $3.21 \pm 0.12$ & $7.88 \pm 0.31$ & $-3.198 \pm 0.097$ & $6.46 \pm 0.35$ & 
$7.91 \pm 0.43$ &   2.73 &  11.98 &  16.71 & S\\
10032840831 & 01423240 & 18.2 & -26.4 & 9.1 &  & $1.378 \pm 0.065$ & 
$0.710 \pm 0.057$ & $6.96 \pm 0.56$ & $-5.73 \pm 0.26$ & $0.782 \pm 0.070$ & 
$7.66 \pm 0.68$ &  16.16 &  65.29 &  87.79 & S\\
10032872401 & 01454801 & 244.9 & -46.3 & 6.5 &  & $3.36 \pm 0.31$ & 
$1.95 \pm 0.22$ & $1.59 \pm 0.17$ & $-2.76 \pm 0.24$ & $3.00 \pm 0.61$ & 
$2.45 \pm 0.50$ &   1.10 &   5.87 &   7.25 & S\\
10033065596 & 01620795 & 246.1 & -22.8 & 2.9 &  & $3.17 \pm 0.12$ & 
$3.47 \pm 0.16$ & $32.6 \pm 1.5$ & $-3.08 \pm 0.12$ & $4.22 \pm 0.33$ & 
$39.6 \pm 3.1$ &  27.22 &  90.86 & 123.39 & S\\
10033165178 & 01706787 & 236.0 & -52.8 & 3.1 &  & $3.43 \pm 0.15$ & 
$3.39 \pm 0.18$ & $4.15 \pm 0.22$ & $-2.90 \pm 0.13$ & $4.51 \pm 0.43$ & 
$5.52 \pm 0.53$ &   1.52 &   7.46 &  11.39 & S\\
10040229862 & 01844268 & 236.3 & -54.0 & 5.0 &  & $3.09 \pm 0.18$ & 
$2.15 \pm 0.15$ & $3.52 \pm 0.25$ & $-3.12 \pm 0.18$ & $2.78 \pm 0.32$ & 
$4.54 \pm 0.53$ &   3.15 &   5.87 &   9.82 & S\\
10041030044 & 02535643 & 270.3 & -23.9 & 13.4 &  & $4.39 \pm 0.59$ & 
$1.75 \pm 0.27$ & $2.14 \pm 0.33$ & $-2.40 \pm 0.27$ & $3.04 \pm 0.94$ & 
$3.7 \pm 1.1$ &   2.52 &   7.26 &  10.29 & S\\
10041112629 & 02604627 & 274.2 & -17.0 & 16.5 &  & $3.24 \pm 0.14$ & 
$3.48 \pm 0.19$ & $5.68 \pm 0.31$ & $-3.10 \pm 0.14$ & $4.35 \pm 0.40$ & 
$7.11 \pm 0.65$ &  11.02 &   4.49 &  16.03 & M\\
\hline
\end{longtable}
\end{center}
\end{tiny}
\end{landscape}

\clearpage
\bibliographystyle{astron}

\bibliography{xrb_catalog}

\hyphenation{Post-Script Sprin-ger}
\begin{thebibliography}{}

\bibitem[\protect\astroncite{{Agostinelli} and {et}}{2003}]{GEANT4}
{Agostinelli}, S. and {et}, a.: 2003,
\newblock {\em Nuclear Instruments and Methods A} {\bf 506}, 250

\bibitem[\protect\astroncite{{Band} et~al.}{1993}]{Band}
{Band}, D., {Matteson}, J., {Ford}, L., {Schaefer}, B., {Palmer}, D.,
  {Teegarden}, B., {Cline}, T., {Briggs}, M., {Paciesas}, W., {Pendleton}, G.,
  {Fishman}, G., {Kouveliotou}, C., {Meegan}, C., {Wilson}, R., and {Lestrade},
  P.: 1993,
\newblock {\em \apj} {\bf 413}, 281

\bibitem[\protect\astroncite{{Belian} et~al.}{1976}]{Belian76}
{Belian}, R.~D., {Conner}, J.~P., and {Evans}, W.~D.: 1976,
\newblock {\em \apjl} {\bf 206}, L135

\bibitem[\protect\astroncite{{Bildsten}}{1998}]{Bildsten98}
{Bildsten}, L.: 1998,
\newblock in {R.~Buccheri, J.~van Paradijs, \& A.~Alpar} (ed.), {\em NATO ASIC
  Proc. 515: The Many Faces of Neutron Stars.}, pp 419--+

\bibitem[\protect\astroncite{Boutloukos et~al.}{2010}]{Boutloukos_2010}
Boutloukos, S., Miller, M.~C., and Lamb, F.~K.: 2010,
\newblock {\em The Astrophysical Journal Letters} {\bf 720(1)}, L15

\bibitem[\protect\astroncite{{Camero-Arranz} et~al.}{2011}]{Arranz_2011}
{Camero-Arranz}, A., {Finger}, M.~H., {Wilson-Hodge}, C., {Jenke}, P.~A.,
  {Steele}, I., {Gutierrez-Soto}, J., and {Coe}, M.~J.: 2011,
\newblock {\em The Astronomer's Telegram} 3166

\bibitem[\protect\astroncite{{Chenevez} et~al.}{2008}]{Chenevez08}
{Chenevez}, J., {Falanga}, M., {Kuuklers}, E., {Brandt}, S., {Lund}, N., and
  {Cumming}, A.: 2008,
\newblock in {\em Proceedings of the 7th INTEGRAL Workshop}, p.~33

\bibitem[\protect\astroncite{{Connaughton} et~al.}{2015}]{Connaughton_2015}
{Connaughton}, V., {Briggs}, M.~S., {Goldstein}, A., {Meegan}, C.~A.,
  {Paciesas}, W.~S., {Preece}, R.~D., {Wilson-Hodge}, C.~A., {Gibby}, M.~H.,
  {Greiner}, J., {Gruber}, D., {Jenke}, P., {Kippen}, R.~M., {Pelassa}, V.,
  {Xiong}, S., {Yu}, H.-F., {Bhat}, P.~N., {Burgess}, J.~M., {Byrne}, D.,
  {Fitzpatrick}, G., {Foley}, S., {Giles}, M.~M., {Guiriec}, S., {van der
  Horst}, A.~J., {von Kienlin}, A., {McBreen}, S., {McGlynn}, S., {Tierney},
  D., and {Zhang}, B.-B.: 2015,
\newblock {\em \apjs} {\bf 216}, 32

\bibitem[\protect\astroncite{{Cornelisse} et~al.}{2002a}]{Cornelisse02b}
{Cornelisse}, R., {Verbunt}, F., {in't Zand}, J.~J.~M., {Kuulkers}, E., and
  {Heise}, J.: 2002a,
\newblock {\em \aap} {\bf 392}, 931

\bibitem[\protect\astroncite{{Cornelisse} et~al.}{2002b}]{Cornelisse02}
{Cornelisse}, R., {Verbunt}, F., {in't Zand}, J.~J.~M., {Kuulkers}, E.,
  {Heise}, J., {Remillard}, R.~A., {Cocchi}, M., {Natalucci}, L., {Bazzano},
  A., and {Ubertini}, P.: 2002b,
\newblock {\em \aap} {\bf 392}, 885

\bibitem[\protect\astroncite{{Cumming} and {Bildsten}}{2000}]{CB00}
{Cumming}, A. and {Bildsten}, L.: 2000,
\newblock {\em \apj} {\bf 544}, 453

\bibitem[\protect\astroncite{Cumming et~al.}{2006}]{Cumming06}
Cumming, A., Macbeth, J., in't Zand, J. J.~M., and Page, D.: 2006,
\newblock {\em \apj} {\bf 646}, 429

\bibitem[\protect\astroncite{{Degenaar} et~al.}{2013}]{Degenaar13b}
{Degenaar}, N., {Miller}, J.~M., {Wijnands}, R., {Altamirano}, D., and
  {Fabian}, A.~C.: 2013,
\newblock {\em \apjl} {\bf 767}, L37

\bibitem[\protect\astroncite{Fujimoto et~al.}{1981}]{Fujimoto81}
Fujimoto, M.~Y., Hanawa, T., and Miyaji, S.: 1981,
\newblock {\em \apj} {\bf 247}, 267

\bibitem[\protect\astroncite{{Galloway} et~al.}{2008}]{Galloway_2008}
{Galloway}, D.~K., {Muno}, M.~P., {Hartman}, J.~M., {Psaltis}, D., and
  {Chakrabarty}, D.: 2008,
\newblock {\em \apjs} {\bf 179}, 360

\bibitem[\protect\astroncite{{Grindlay} et~al.}{1976}]{Grindlay76}
{Grindlay}, J., {Gursky}, H., {Schnopper}, H., {Parsignault}, D.~R., {Heise},
  J., {Brinkman}, A.~C., and {Schrijver}, J.: 1976,
\newblock {\em \apjl} {\bf 205}, L127

\bibitem[\protect\astroncite{{Hakala} et~al.}{2005}]{Hakala05}
{Hakala}, P., {Ramsay}, G., {Muhli}, P., {Charles}, P., {Hannikainen}, D.,
  {Mukai}, K., and {Vilhu}, O.: 2005,
\newblock {\em \mnras} {\bf 356}, 1133

\bibitem[\protect\astroncite{{Harris}}{1996}]{Harris96}
{Harris}, W.~E.: 1996,
\newblock {\em \aj} {\bf 112}, 1487

\bibitem[\protect\astroncite{{in't Zand}}{2005}]{intZand05}
{in't Zand}, J.~J.~M.: 2005,
\newblock {\em \aap} {\bf 441}, L1

\bibitem[\protect\astroncite{{in't Zand} et~al.}{2008}]{intZand08}
{in't Zand}, J.~J.~M., {Bassa}, C.~G., {Jonker}, P.~G., {Keek}, L., {Verbunt},
  F., {M{\'e}ndez}, M., and {Markwardt}, C.~B.: 2008,
\newblock {\em \aap} {\bf 485}, 183

\bibitem[\protect\astroncite{{in't Zand} et~al.}{2007}]{intZand07}
{in't Zand}, J.~J.~M., {Jonker}, P.~G., and {Markwardt}, C.~B.: 2007,
\newblock {\em \aap} {\bf 465}, 953

\bibitem[\protect\astroncite{{Kuulkers} et~al.}{2003}]{Kuulkers_2003}
{Kuulkers}, E., {den Hartog}, P.~R., {in't Zand}, J.~J.~M., {Verbunt},
  F.~W.~M., {Harris}, W.~E., and {Cocchi}, M.: 2003,
\newblock {\em \aap} {\bf 399}, 663

\bibitem[\protect\astroncite{{Kuulkers} et~al.}{2010}]{Kuulkers_2010}
{Kuulkers}, E., {in't Zand}, J.~J.~M., {Atteia}, J.-L., {Levine}, A.~M.,
  {Brandt}, S., {Smith}, D.~A., {Linares}, M., {Falanga}, M.,
  {S{\'a}nchez-Fern{\'a}ndez}, C., {Markwardt}, C.~B., {Strohmayer}, T.~E.,
  {Cumming}, A., and {Suzuki}, M.: 2010,
\newblock {\em \aap} {\bf 514}, A65

\bibitem[\protect\astroncite{{Lewin} et~al.}{1993}]{Lewin93}
{Lewin}, W.~H.~G., {van Paradijs}, J., and {Taam}, R.~E.: 1993,
\newblock {\em \ssr} {\bf 62}, 223

\bibitem[\protect\astroncite{{Linares} et~al.}{2012}]{Linares12b}
{Linares}, M., {Connaughton}, V., {Jenke}, P., {van der Horst}, A.~J.,
  {Camero-Arranz}, A., {Kouveliotou}, C., {Chakrabarty}, D., {Beklen}, E.,
  {Bhat}, P.~N., {Briggs}, M.~S., {Finger}, M., {Paciesas}, W.~S., {Preece},
  R., {von Kienlin}, A., and {Wilson-Hodge}, C.~A.: 2012,
\newblock {\em \apj} {\bf 760}, 133

\bibitem[\protect\astroncite{{Matsuoka} et~al.}{2009}]{Maxi_2009}
{Matsuoka}, M., {Kawasaki}, K., {Ueno}, S., {Tomida}, H., {Kohama}, M.,
  {Suzuki}, M., {Adachi}, Y., {Ishikawa}, M., {Mihara}, T., {Sugizaki}, M.,
  {Isobe}, N., {Nakagawa}, Y., {Tsunemi}, H., {Miyata}, E., {Kawai}, N.,
  {Kataoka}, J., {Morii}, M., {Yoshida}, A., {Negoro}, H., {Nakajima}, M.,
  {Ueda}, Y., {Chujo}, H., {Yamaoka}, K., {Yamazaki}, O., {Nakahira}, S.,
  {You}, T., {Ishiwata}, R., {Miyoshi}, S., {Eguchi}, S., {Hiroi}, K.,
  {Katayama}, H., and {Ebisawa}, K.: 2009,
\newblock {\em \pasj} {\bf 61}, 999

\bibitem[\protect\astroncite{{Meegan} et~al.}{2009}]{Meegan2009}
{Meegan}, C., {Lichti}, G., {Bhat}, P.~N., {Bissaldi}, E., {Briggs}, M.~S.,
  {Connaughton}, V., {Diehl}, R., {Fishman}, G., {Greiner}, J., {Hoover},
  A.~S., {van der Horst}, A.~J., {von Kienlin}, A., {Kippen}, R.~M.,
  {Kouveliotou}, C., {McBreen}, S., {Paciesas}, W.~S., {Preece}, R., {Steinle},
  H., {Wallace}, M.~S., {Wilson}, R.~B., and {Wilson-Hodge}, C.: 2009,
\newblock {\em \apj} {\bf 702}, 791

\bibitem[\protect\astroncite{{Murakami} et~al.}{1983}]{Murakami83}
{Murakami}, T., {Inoue}, H., {Koyama}, K., {Makishima}, K., {Matsuoka}, M.,
  {Oda}, M., {Ogawara}, Y., {Ohashi}, T., {Makino}, F., {Shibazaki}, N.,
  {Tanaka}, Y., {Hayakawa}, S., {Kunieda}, H., {Nagase}, F., {Masai}, K.,
  {Tawara}, Y., {Miyamoto}, S., {Tsunemi}, H., {Yamashita}, K., and {Kondo},
  I.: 1983,
\newblock {\em \pasj} {\bf 35}, 531

\bibitem[\protect\astroncite{{Nelemans} et~al.}{2004}]{Nelemans04}
{Nelemans}, G., {Jonker}, P.~G., {Marsh}, T.~R., and {van der Klis}, M.: 2004,
\newblock {\em \mnras} {\bf 348}, L7

\bibitem[\protect\astroncite{{Ott} et~al.}{2011}]{Ott_2011}
{Ott}, C.~D., {Reisswig}, C., {Schnetter}, E., {O'Connor}, E., {Sperhake}, U.,
  {L{\"o}ffler}, F., {Diener}, P., {Abdikamalov}, E., {Hawke}, I., and
  {Burrows}, A.: 2011,
\newblock {\em Physical Review Letters} {\bf 106(16)}, 161103

\bibitem[\protect\astroncite{{Serino} et~al.}{2015}]{Serino15}
{Serino}, M., {Shidatsu}, M., {Ueda}, Y., {Matsuoka}, M., {Negoro}, H.,
  {Yamaoka}, K., {Kennea}, J.~A., {Fukushima}, K., and {Nagayama}, T.: 2015,
\newblock {\em \pasj} {\bf 67}, 30

\bibitem[\protect\astroncite{{Swank} et~al.}{1977}]{Swank77}
{Swank}, J.~H., {Becker}, R.~H., {Boldt}, E.~A., {Holt}, S.~S., {Pravdo},
  S.~H., and {Serlemitsos}, P.~J.: 1977,
\newblock {\em \apjl} {\bf 212}, L73

\bibitem[\protect\astroncite{{Weinberg} et~al.}{2006}]{Weinberg06}
{Weinberg}, N.~N., {Bildsten}, L., and {Schatz}, H.: 2006,
\newblock {\em \apj} {\bf 639}, 1018

\end{thebibliography}
\clearpage

\appendix
\section{aFXPs Light Curves}  \label{lc:aFXPs}
See http://gammaray.msfc.nasa.gov/gbm/science/xrb.html
\section{uGRBs Light Curves} \label{lc:uGRBs}
See http://gammaray.msfc.nasa.gov/gbm/science/xrb.html
\section{tXRBs Light Curves} \label{lc:tXRBs}
See http://gammaray.msfc.nasa.gov/gbm/science/xrb.html
\end{document}